\documentclass[aps,prd,reprint,groupedaddress]{revtex4-1}

\usepackage{color}
\usepackage{graphicx}
\usepackage{amssymb}
\usepackage{multirow}
\usepackage{array}
\usepackage{subfigure}

\usepackage{aas_macros}

\begin{document}

\title{Astrophysical searches for a hidden-photon signal in the radio regime}

\author{Andrei\,P.~Lobanov}
\email[]{alobanov@mpifr-bonn.mpg.de}
\affiliation{Max-Planck-Institut f\"ur Radioastronomie, Auf dem
  H\"ugel 69, 53121 Bonn, Germany}
\altaffiliation{Visiting Scientist,
  University of Hamburg / Deutsches Elektronen Synchrotron (DESY)
  Forschungszentrum}
\author{Hannes-S. Zechlin}
\email[]{hzechlin@physik.uni-hamburg.de}
\author{Dieter Horns}
\email[]{dieter.horns@physik.uni-hamburg.de}
\affiliation{University of Hamburg, Institut f\"ur Experimentalphysik, Luruper Chaussee 149,
  D-22761 Hamburg, Germany}

\date{5 February 2013}

\begin{abstract}
  Common extensions of the Standard Model of particle physics predict
  the existence of a ``hidden'' sector that comprises particles with a
  vanishing or very weak coupling to particles of the Standard Model
  (visible sector). For very light ($m < 10^{-14}$~eV) hidden U(1)
  gauge bosons (hidden photons), broad-band radio spectra of compact
  radio sources could be modified due to weak kinetic mixing with
  radio photons. Here, search methods are developed and their
  sensitivity discussed, with specific emphasis on the effect of the
  coherence length of the signal, instrumental bandwidth, and spectral
  resolution. We conclude that radio observations in the frequency
  range of 0.03--1400\,GHz probe kinetic mixing of $\sim\!10^{-3}$ of
  hidden photons with masses down to $\sim\!10^{-17}$\,eV. Prospects
  for improving the sensitivity with future radio astronomical
  facilities as well as by stacking data from multiple objects are
  discussed.
\end{abstract}



\maketitle

\section{Introduction}
Finding experimental evidence for physics beyond the Standard Model
(SM) of particle physics is one of the pinnacles of present-day
physical research, embracing both extensive laboratory studies and
indirect (primarily) astrophysical measurements made across a very
broad range of energies. Most of the present-day SM extensions
  into a more generic, unified scenario predict existence of a class
  of particles only weakly interacting with the normal matter. These
  particles are typically divided into two broad classes defined by
  the particle mass, with weakly interacting massive particles (WIMP)
  acting at a mass scale of $\mathcal{O}(100)$\,GeV
  \cite{jungman1996,bertone2005,bertone2010} and ultralight weakly
  interacting sub-eV particles (WISP) whose masses can be $\ll 1$\,eV
  \cite{okun1982,georgi1984,bergstrom2000,ahlers2008,jaeckel2010,ringwald2012}.

The existence of ultralight particles has been argued to be at least
theoretically plausible in a number of different scenarios, including
additional pseudo-scalar (axions and axion-like particles (ALP),
$\phi$, \cite{jaeckel2007,raffelt1988,zavattini2006}) as well as
vector fields (U(1) hidden photons, $\gamma_\mathrm{s}$,
\cite{okun1982,georgi1984,ahlers2007,ahlers2008,jaeckel2010}). In
either scenario, the prevailing non-baryonic matter could be explained
by these ultralight fields \cite{redondo2009,arias2012}.

Hidden photons arise in low-energy extensions of the SM which leave
all the SM fields uncharged under the additional $U(1)$ gauge group.
Interaction between hidden photons and massive SM particles is
expected to be suppressed by the particle masses. However,
it can be manifested by kinetic mixing with normal photons
\cite{okun1982,holdom1986,ahlers2007}. In low-energy SM extensions
with hidden photons, the kinetic mixing is expressed by the effective
Lagrangian describing two-photon interactions ${\cal L} = {\cal
  L}_\mathrm{SM} + {\cal L}_\mathrm{h} + {\cal L}_{\chi}$
\cite[\emph{cf.}][]{jaeckel2008,redondo2008a,redondo2009}, with ${\cal
  L}_\mathrm{SM}$ denoting the Maxwell-Lagrangian for the SM photon
field, the ${\cal L}_\mathrm{h}$ term describing the Proca-Lagrangian
for the massive hidden-photon field, and ${\cal L}_{\chi}$
representing a gauge-invariant kinetic mixing term. The kinetic mixing
term induces photon oscillations between the massless ``normal'' state
($\gamma$) and a non-zero mass ``hidden'' state
($\gamma_\mathrm{s}$). In this hidden state, photons acquire a
non-vanishing mass and propagate on time-like geodesics, without any
interaction with normal matter. The physical properties of a hidden
photon can be completely described by its mass $m_{\gamma_\mathrm{s}}$
and the kinetic mixing with a SM photon (expressed by the mixing angle
$\chi$).  Theoretical predictions for $\chi$ fall in the broad range
between $10^{-16}$ and $10^{-2}$
\cite{holdom1986,dienes1997,abel2008a,abel2008b,goodsell2009,cicoli2011}.

Accelerator experiments are generally optimized to search for new
heavy particles such as WIMPs, and therefore of limited sensitivity
and mass reach for WISPs. Hence, the potential discovery of ultralight
particles requires high-precision experiments for which
non-accelerator setups often appear more promising
\cite{asztalos2010,sikivie2010,baker2012}.

Pioneering work on the phenomenology of hidden photons has
  already focussed on astrophysical implications of low-mass hidden
  photons \cite{okun1982,georgi1984}. Subsequently, evidence for the
$\gamma$--$\gamma_\mathrm{s}$ oscillations has been searched for in a
number of laboratory
\cite{afanasiev2010,ehret2010,cadamuro2010,wagner2010,andreas2012,betz2012}
and astrophysical experiments
\cite{zechlin2008,redondo2008a,mirizzi2009a,arias2010,redondo2010a,jaeckel2010,cadamuro2010},
focusing in particular on ``light shining through the wall'' (LSW,
\cite{redondo2011}) experiments such as ALPS \cite{ehret2010} and
searches for hidden photons from the Sun (SHIPS;
\cite{cadamuro2010,schwarz2011}). The non-detection of hidden-photon
signals has so far yielded strong bounds on the kinetic mixing
parameter $\chi$ for a broad range of hidden-photon masses \cite[][and
  references therein]{ahlers2008,jaeckel2010,arias2012}. The mass
range which has been currently probed extends down to
$m_{\gamma_\mathrm{s}} = 2\times 10^{-14}$\,eV, with the lowest
hidden-photon masses probed by the WMAP CMB measurements in the radio
domain at frequencies above 22\,GHz \cite{mirizzi2009a}. Below
$10^{-14}$\,eV, only weak limits of $\chi \approx 10^{-2}$
\cite{ahlers2008} have been obtained from analysis of early
measurements of magnetic fields around Earth and Jupiter
\cite{goldhaber1971}, and no limits are reported for
$m_{\gamma_\mathrm{s}} \lesssim 5\times 10^{-16}$\,eV.

Radio observations at frequencies below 22\,GHz offer an excellent (if
not unique) tool for placing bounds on the mixing angle $\chi$ for
$m_{\gamma_\mathrm{s}} < 10^{-14}$\,eV. Initial bounds on $\chi$ can
be obtained from existing radio data on compact, weakly variable
objects with well-known radio spectra (such as young supernova
remnants (SNR), planetary nebula, and steep spectrum radio sources
typically used for the absolute flux density calibration of radio
telescopes). With this approach, one can reasonably expect to reach
$\chi \lesssim 0.01$. Propagation through a refractive medium (in
which an SM photon also acquires an effective mass $m_{\gamma}$) can
strongly affect this limit, improving it substantially near the
resonance condition $m_{\gamma_\mathrm{s}} = m_{\gamma}$, and
suppressing the hidden photon conversion at $m_{\gamma}\gg
m_{\gamma_\mathrm{s}}$ \cite{mirizzi2009a}. The latter effect can
become important at $m_{\gamma_\mathrm{s}} \lesssim 10^{-15}$\,eV.

Placing better bounds on $\chi$ can now be achieved by using the
expanded capabilities of existing radio telescopes (utilizing the
upgraded broad-band coverage and spectral resolution of the Effelsberg
100-meter antenna and the Karl Jansky Very Large Array (JVLA) in the
0.3--40\,GHz frequency range), by extending the measurements both to
lower frequencies (0.03--0.3\,GHz) covered by the Low Frequency Array
(LOFAR) and to submillimeter wavelengths probed by Atacama Large
Millimeter Array (ALMA), and by employing the superb brightness
sensitivity ($\sim\!1\,\mu$Jy \footnote{1\,Jy =
  $10^{-26}$\,J\,m$^{-2}$\,s$^{-1}$\,Hz$^{-1}$}) of the
SKA\footnote{Square Kilometer Array, a next generation radio telescope
  that will provide about a two orders of magnitude improvement in
  imaging sensitivity and surveying speed for radio observations in
  the 0.3--20\,GHz frequency range; {\tt
    http://www.skatelescope.org}.}  and its precursors,
MeerKAT\footnote{{\tt http://www.ska.ac.za/meerkat}} and
ASKAP\footnote{{\tt http://www.atnf.csiro.au/SKA}}.

The existing data for the primary absolute flux density calibrators in
the radio regime (such as Cas~A, Tau~A, and Cyg~A) \cite{baars1977}
feature a few dozens of absolute flux density measurements made in the
0.01--30\,GHz spectral range and reaching a $\sim\!3$--$5\%$ accuracy.
These data should enable placing a bound of $\chi \approx 0.02$, from
measurements of the r.m.s. of deviations from a canonical source
spectrum. If the spectral resolution is sufficiently high to assess
the periodicity in the oscillation signal (particularly at the lower
end of the spectrum), both the limits on $\chi$ and the range of
photon mass studied can be improved. Further improvements of the
bounds on $\chi$ can be achieved by stacking the signal from a number
of objects, under the condition that the observations of different
objects are sensitive to the same range of the hidden-photon mass.

In this paper, a methodology and prospects for detection of the hidden
photon signal in the radio regime are considered. The basic physics of
the $\gamma$--$\gamma_\mathrm{s}$ oscillation and the propagation of
the hidden-photon signal are described in
Section~\ref{sc:phys}. Methods for the detection of the oscillation
signal with different instruments and targets are discussed in
Section~\ref{sc:detection} and potentials of these studies are
discussed in Section \ref{sc:disc}.

\section{Photon oscillations in the radio regime}
\label{sc:phys}

For a photon field, $A_\mu$, and a hidden photon field, $B_{\mu}$,
the kinetic mixing term is given by \cite{jaeckel2008}
\begin{equation}
{\cal L}_\chi = \frac{\sin\chi}{2}A_{\mu\nu} B^{\mu\nu} +
\frac{\cos^{2}\chi}{2}m_{\gamma_\mathrm{s}}^2 B_{\mu} B^{\mu}\,,
\end{equation}
where $A_{\mu\nu}$ and $B_{\mu\nu}$ are the respective field-strength
tensors. The non-zero kinetic mixing angle $\chi$ implies a mismatch
between the interaction and propagation eigenstates, which induces
oscillation between the two states (with a close analogy to the
neutrino oscillation effect \cite{kuo1989}). The last term accounts
for massive hidden photons via Higgs or St\"uckelberg mechanisms,
where the former case suffers from additional constraints
\cite{ahlers2008}. The probability of the
$\gamma\rightarrow\gamma_\mathrm{s}$ conversion after propagating a
distance $L$ in vacuum is then given by
\begin{equation} \label{eq:osc_prob}
P_{\gamma\rightarrow\gamma_\mathrm{s}}(L) = a_\chi
\sin^2\left(\frac{m_{\gamma_\mathrm{s}}^2}{4 E} L\right) = a_\chi
\sin^2\left(\frac{m_{\gamma_\mathrm{s}}^2}{8 \pi \nu} L\right),
\end{equation}
where all quantities are expressed in natural units, and $E$ and $\nu$
are the energy and frequency of the normal photon
\cite{ahlers2007,jaeckel2008,zechlin2008}. It should be noted that the
validity of Eq.~\ref{eq:osc_prob} is restricted to the case of
$m_{\gamma_\mathrm{s}} \ll 2\pi\nu$, which is fulfilled for all
considerations in this paper.

The first term of Eq.~\ref{eq:osc_prob}
describes the amplitude of the oscillation, $a_\chi = \sin^2
(2\chi) \approx 4 \chi^2$ (for $\chi \ll 1$), that can be
identified as a periodic signal over a range of distances $L$ or
wavelengths $\lambda = 1/\nu$. The second term,
\begin{equation}
\varphi_{\gamma_\mathrm{s}}(\nu) =
\frac{m_{\gamma_\mathrm{s}}^2 L}{8\pi\nu} = 9.45
\left(\frac{m_{\gamma_\mathrm{s}}}{10^{-15}\,\mathrm{eV}}\right)^2
\left(\frac{L}{\mathrm{pc}}\right)
\left(\frac{\nu}{\mathrm{MHz}}\right)^{-1},
\end{equation}
gives the periodic signature of the oscillation. The oscillation
signal affects the broad-band spectrum, $F(\nu)$, of an astrophysical
source, which results in a received spectrum
$F_{\gamma_\mathrm{s}}(\nu) =
F(\nu)\,(1-P_{\gamma\rightarrow\gamma_\mathrm{s}})$.

The resulting spectrum $F_{\gamma_\mathrm{s}}(\nu)$ will have local
minima and maxima at the frequencies $\nu_{\mathrm{min},i} =
\nu_\star/(2i-1)$ and $\nu_{\mathrm{max},i} =
\nu_\star/(2i)$, $i \in \mathbb{N}$, where
\begin{equation}
\nu_\star = \frac{m_{\gamma_\mathrm{s}}^2 L}{4\pi^2} = 6.02\,
\left(\frac{m_{\gamma_\mathrm{s}}}{10^{-15}\,\mathrm{eV}}\right)^2
\left(\frac{L}{\mathrm{pc}}\right)\,\, \mathrm{MHz}
\end{equation}
is the frequency of the first (highest frequency) minimum
obtained with $i=1$. The frequency, $\nu_\mathrm{max,1}$, of
the first maximum defines the characteristic wavelength,
\begin{equation}
\lambda_\star = \frac{8\pi^2}{m_{\gamma_\mathrm{s}}^2 L} = 99.64\,
\left(\frac{m_{\gamma_\mathrm{s}}}{10^{-15}\,\mathrm{eV}}\right)^{-2}
\left(\frac{L}{\mathrm{pc}}\right)^{-1}\,\, \mathrm{m}\,,
\end{equation}
of the periodic modulation induced by the hidden-photon signal on a
broad-band spectrum $F(\lambda)$ in the wavelength domain.

\subsection{Oscillation and coherence lengths}\label{ssec:osc_coh}

The oscillation length of the hidden-photon signal is 
set by the condition $\varphi_{\gamma_\mathrm{s}}(L,\nu) = \pi$,
\begin{equation} \label{eq:losc}
L_\mathrm{osc} = 0.33\,\left(\frac{\nu}{\mathrm{MHz}}\right)
\left(\frac{m_{\gamma_\mathrm{s}}}{10^{-15}\,\mathrm{eV}}\right)^{-2}\mathrm{pc}\,.
\end{equation}
However, decoherence effects in the photon and hidden-photon mass
eigenstates may arise during propagation, owing to the finite mass,
$m_{\gamma_\mathrm{s}}$, of the latter. An accurate quantum mechanical
treatment of the oscillation probability $P_{\gamma \rightarrow
  \gamma_\mathrm{s}}$ using wave packets (Eq. \ref{eq:osc_prob} has
been derived assuming plane waves) yields an upper bound on the
accessible distance range \cite{nussinov1976,giunti1998,zechlin2008},
$L_\mathrm{coh} = 4\sqrt{2}\sigma_x E^2/m_{\gamma_\mathrm{s}}^2$,
where $\sigma_x^2 = \sigma_{x,\mathrm{P}}^2 + \sigma_{x,\mathrm{D}}^2$
denote the quantum mechanical uncertainties of the production and
detection processes, respectively. The non-thermal radio emission of
compact radio sources is produced by synchrotron radiation, with the
mean energy loss path of synchrotron emitting electrons
$\sigma_{x,\mathrm{P}} = \Delta \tau$. Note that the mean energy loss
path is of the same order of magnitude as the gyro radius of the
relativistic electrons under consideration. A reasonable estimate of
the cooling time $\Delta \tau = 2\pi\nu_c/(-\mathrm{d}E/\mathrm{d}t)$
can be obtained at the critical frequency $\nu_c$ of synchrotron
radiation (averaging over the pitch angle) \cite{meyer2010},
eventually yielding
\begin{equation}\label{eq:dmax}
L_\mathrm{coh} = 19.84\,\left(\frac{\nu}{\mathrm{MHz}}\right)^2
\left(\frac{m_{\gamma_\mathrm{s}}}{10^{-15}\,\mathrm{eV}}\right)^{-2} 
\left(\frac{B}{\mathrm{mG}}\right)^{-1}\,\mathrm{kpc}\,,
\end{equation}
where $B$ denotes the magnetic field inside the considered source. The
$\gamma$--$\gamma_\mathrm{s}$ oscillations can therefore be probed in
vacuum at any distance $L$ fulfilling the condition $L_\mathrm{osc}
\leq L \leq L_\mathrm{coh}$.

\begin{table*}[t]
\centering
\caption{Lower and upper bounds on the mass range of hidden photons,
  $m_{\gamma_\mathrm{s}}^\ell$ and
  $m_{\gamma_\mathrm{s}}^\mathrm{u}$, respectively, detectable with
  radio data ($\nu > 1\,\mathrm{GHz}$) of SNR and AGN. For
  $m_{\gamma_\mathrm{s}} < m_{\gamma_\mathrm{s}}^\ell$, no
  $\gamma$--$\gamma_\mathrm{s}$ oscillations will arise; for
  $m_{\gamma_\mathrm{s}} > m_{\gamma_\mathrm{s}}^\mathrm{u}$,
  decoherence effects yield a freeze out of the oscillation
  signal. The columns $B$ and $\Delta L$ list the assumed magnetic
  fields and distance ranges.} \label{tab:mass_bounds}
\begin{ruledtabular}
\begin{tabular}{lcc>{\centering}p{5cm}<{\centering}c}
Source class & $B$\,[mG] & $\Delta L$ & $m_{\gamma_\mathrm{s}}^\ell$\,[eV] & $m_{\gamma_\mathrm{s}}^\mathrm{u}$\,[eV] \\
\hline
SNR & 0.1 & 1--10\,kpc & $2\times 10^{-16}\!-\!6\times 10^{-16}$ & $4\times 10^{-12}\!-\!1\times 10^{-11}$ \\
AGN: radio lobes & 1 & 0.02--3\,Gpc & $3\times 10^{-19}\!-\!4\times 10^{-18}$ & 
$3\times 10^{-15}\!-\!3\times 10^{-14}$ \\
AGN: nuclear regions & $10^3$ & 0.02--3\,Gpc & $3\times 10^{-19}\!-\!4\times 10^{-18}$ & $10^{-16}\!-\!10^{-15}$ \\
\end{tabular}
\end{ruledtabular}
\end{table*}

Together with this condition, Eqs.~\ref{eq:losc} and \ref{eq:dmax}
imply effective lower and upper bounds on the hidden-photon mass that
can be probed with radio data of a compact synchrotron emitting source
at distance $L_0$ and above a given frequency
$\nu_0$. Table~\ref{tab:mass_bounds} lists corresponding bounds
considering three distinct types of radio sources, namely nearby
Galactic supernova remnants (SNR), distant radio-emitting lobes of
active galactic nuclei (AGN), and AGN cores at cosmological
distances. It demonstrates that the oscillation length and decoherence
effects should enable effective radio measurements for a range of
hidden-photon masses between $\sim\! 10^{-19}\,\mathrm{eV}$ and
$\sim\! 10^{-11}$\,eV.

\subsection{Propagation through refractive media}
\label{sc:refr}

Photon propagation through a medium with refractive index $n$ can be
described by introducing an effective photon mass $m_{\gamma}$ to the
Lagrangian ${\cal L}$. This operation affects the kinetic mixing term,
and the resulting effective mixing angle $\chi_\mathrm{r}$ depends on
the mass ratio $\xi = m_\gamma^2/m_{\gamma_\mathrm{s}}^2$ so that
\cite{jaeckel2008,redondo2008a,mirizzi2009a}
\begin{equation}
\sin 2\chi_\mathrm{r} = \frac{\sin 2\chi}{\sqrt{\sin^2 2\chi + 
(\cos 2\chi-\xi)^2}}\,.
\end{equation}
For small effective photon masses, $\xi \ll 1$, the
$\gamma$--$\gamma_\mathrm{s}$ oscillations approach the vacuum regime,
with $\chi_\mathrm{r} \rightarrow \chi$. A resonance with the maximum
amplitude $\chi_\mathrm{r} = \pi/4$ of the oscillations is reached at
the resonant mass ratio $\xi = \cos 2\chi$. For higher effective
photon masses, the oscillations are rapidly damped (medium
suppression), with $\chi_\mathrm{r} \rightarrow \pi/2$ for $\xi \gg
1$. The condition $\sin 2\chi_\mathrm{r} \ge \sin 2\chi$ implies that
hidden photons with a given $m_{\gamma_\mathrm{s}}$ and $\chi$ can be
detected in a medium with the effective photon mass $m_\gamma^2 \le
2\,m_{\gamma_\mathrm{s}}^2 \cos 2\chi$, which can be approximated with
$m_\gamma \lesssim \sqrt{2}\,m_{\gamma_\mathrm{s}}$ for $\chi \ll 1$.

For photon propagation in the interstellar (ISM) and intergalactic
(IGM) medium, the dominant factor is scattering off free electrons and
neutral atoms (with the medium described by the electron and proton
number densities, $n_\mathrm{e}$ and $n_\mathrm{p}$). In this case,
the effective photon mass, $m_\gamma^2 \approx \omega_\mathrm{P}^2 - 2
\omega^2 (n-1)_\mathrm{medium}$ \cite{mirizzi2009a}, depends on the
photon frequency $\omega$, the plasma frequency $\omega_\mathrm{P}^2 =
4\pi \alpha n_\mathrm{e}/m_\mathrm{e}$, and the refraction index
$(n-1)_\mathrm{medium}$ of the medium.

For Galactic objects and extragalactic objects at small redshifts ($z
<1$), the medium can be assumed strongly ionized (with the ionized
fraction of hydrogen $X_\mathrm{e} = n_\mathrm{e}/n_\mathrm{p}
\rightarrow 1$). Contributions from helium and heavier elements can be
neglected. This yields \cite{mirizzi2009a}
\begin{eqnarray}
m_{\gamma}^2 & \approx & \omega_\mathrm{P}^2 - 2 \omega^2
  (n-1)_\mathrm{medium} \\ \nonumber
 & \simeq & 1.4 \times 10^{-21}
\left(\frac{n_\mathrm{p}}{\mathrm{cm}^{-3}}\right) \\ \nonumber
 &  & \times \left[X_\mathrm{e} - 1.2 \times 10^{-19}
  \left(\frac{\nu}{\mathrm{MHz}}\right)^2 (1-X_\mathrm{e})
  \right]\mathrm{eV}^2\,.
\end{eqnarray}
For observations in the radio band, $m_{\gamma} \approx
\omega_\mathrm{P} \simeq 3.7 \times 10^{-11}\,
(n_\mathrm{e}/\mathrm{cm}^{-3})^{1/2}\,\mathrm{eV}$ provides a good
estimate of the effective photon mass under the assumption of $X_e
\approx 1$. This limits, formally, the hidden-photon mass that can be
probed in Galactic and extragalactic objects to $\sim 10^{-13}$\,eV
and $\sim 10^{-14}$\,eV, respectively (for generic assumptions for the
average densities of $\tilde{n}_\mathrm{ISM} \approx
10^{-4}$\,cm$^{-3}$ and $\tilde{n}_\mathrm{IGM} \approx
10^{-6}$\,cm$^{-3}$ \cite{hinshaw2009}). However, both the ISM and the
IGM are inhomogeneous, with variations of the density exceeding two to
three orders of magnitude \cite{gottloeber2003,pynzar2008}. Hence,
generation and propagation of the hidden-photon signal depend on the
line-of-sight (LOS) properties of the medium and its inhomogeneities.

\subsection{Effects of inhomogeneous media}
\label{sc:inhom}

In an inhomogeneous medium, the minimum value of
$m_{\gamma_\mathrm{s}}$ without medium damping is limited by the
electron density of underdense regions with $n_\mathrm{e} \ll
\tilde{n}_\mathrm{e}$. The traversed underdense region needs to be
sufficiently extended to affect the propagation, such that the LOS
pathlength, $l_\mathrm{los}$, in these regions is $l_\mathrm{los} \gg
L_\mathrm{osc}$ (for observations in the radio domain at $\nu\gtrsim
100$\,MHz, this effectively limits $m_{\gamma_\mathrm{s}} \gtrsim
10^{-16}$\,eV for Galactic objects, while permitting probing hidden
photons with $m_{\gamma_\mathrm{s}} \gtrsim 10^{-19}$\,eV with
extragalactic targets, see Table~\ref{tab:mass_bounds}).  Given
  steep density gradients typically found at the edges of voids
  \cite[{\em cf., \rm}][]{pan2012}, one can reasonably assume that
  propagation through voids would not significantly distort the
  oscillation pattern. The resulting spectral pattern after
propagating through underdense regions remains ``frozen'' during
subsequent propagation through denser regions (where
$m_{\gamma_\mathrm{s}} \gtrsim 2.6 \times 10^{-11}
(n_\mathrm{e}/\mathrm{cm}^{-3})^{1/2}$), as both direct and reverse
photon conversions are suppressed there.  Therefore, for Galactic
objects, the lowest detectable hidden-photon mass would be achieved
for photon beams propagating between the Galactic arms, while
propagation through cosmic voids would set the lower limit on the
hidden-photon mass that can be detected in the broad-band spectra of
extragalactic targets.

A density of free electrons $n_\mathrm{e,loc} \approx
0.005$--$0.01$\,cm$^{-3}$ is measured in the local ISM
\cite{taylor1993,cordes2002}, and there is ample evidence for
$n_\mathrm{e}$ to vary strongly across the Galaxy
\cite{taylor1993,cordes2002}, with $n_\mathrm{e} \gtrsim
10\,\mathrm{cm}^{-3}$ near the Galactic center, $n_\mathrm{e} \sim
10^{-2}\,\mathrm{cm}^{-3}$ in the spiral arms, and $n_\mathrm{e} \ll
10^{-4}\,\mathrm{cm}^{-3}$ above the Galactic disk.  In mini ``void''
regions of $\sim 1$\,kpc in extent and located between the spiral arms
\cite{cordes2002}, $n_\mathrm{e} \lesssim 10^{-6}\,\mathrm{cm}^{-3}$
can be found \cite{pynzar2008}, which is similar to the values
typically measured in the IGM. Hence, detectability of hidden-photon
oscillation in Galactic sources should depend strongly on the LOS to a
specific target, with likely $m_{\gamma_\mathrm{s}} \gtrsim
10^{-12}\,\mathrm{eV}$ detectable for LOS not crossing the spiral
arms, while $m_{\gamma_\mathrm{s}} \gtrsim 10^{-14}\,\mathrm{eV}$ may
still be detectable for objects at high galactic latitudes, and the
LOS crossing inter-arm plasma and the local ``voids''.

The electron and proton densities in different structural components
of the IGM can be estimated from observations, i.e., IRAS data
\cite{plionis2002} and SDSS data \cite{pan2012}, as well as detailed
numerical simulations of large-scale structures
\cite{kravtsov2002,gottloeber2003,cen2005}. The SDSS data indicates that the
voids have a volume filling factor of 0.62 and a median size of
$17\,h^{-1}$ Mpc, and Ref. \cite{gottloeber2003} finds the baryonic
matter density, $\Omega_\mathrm{b,void} \approx (0.045 \pm 0.015)\,
\Omega_\mathrm{b}$, where $h =
H/(100\,\mathrm{km}\,\mathrm{s}^{-1}\,\mathrm{Mpc}^{-1})$ is the
dimensionless Hubble constant and $\Omega_\mathrm{b} =0.046 \pm 0.002$
is the average baryon density in the Universe
\cite{komatsu2011}. Table~\ref{tb:igm} summarizes the mass fraction,
$m/m_\mathrm{b}$, the volume fraction, $V/V_\mathrm{c}$, and the
relative density, $\rho/\rho_\mathrm{b}$, of different IGM components
(the warm, warm-hot ionized (WHIM), hot, and void components, measured
with respect to the total baryon mass $m_\mathrm{b}$, density
$\rho_\mathrm{b}$, and the comoving volume $V_\mathrm{c}$). Based on
these values, one can estimate the electron density in the individual
IGM components
\begin{equation}
n_\mathrm{e,medium} = X_\mathrm{e} \frac{\rho_\mathrm{c} \Omega_\mathrm{b}}{m_\mathrm{p} +
 X_\mathrm{e} m_\mathrm{e}}  \frac{\rho_\mathrm{medium}}{\rho_\mathrm{b}}\,,
\end{equation}
where $\rho_\mathrm{c} = 3H^2/(8\pi G)$ is the critical density of the
Universe, $m_\mathrm{p}$ and $m_\mathrm{e}$ are the proton and
electron masses, and $G$ denotes the gravitational constant. This
yields $n_\mathrm{e}(\rho_\mathrm{b}) = 2.5 \times
10^{-7}\,\mathrm{cm}^{-3}$ and enables calculating $n_\mathrm{e}$ and
respective limits on $m_{\gamma_\mathrm{s}}$ for different baryonic
matter components as listed in Table~\ref{tb:igm} (compiled from the
results reported in
\cite{plionis2002,pan2012,kravtsov2002,gottloeber2003,cen2005}).

\begin{table}[ht]
\caption{Baryonic matter components and hidden-photon
  propagation. Each IGM component is described by the temperature $T$,
  mass fraction $m/m_\mathrm{b}$, comoving volume fraction
  $V/V_\mathrm{c}$, local density relative to the average baryon
  density $\rho/\rho_\mathrm{b}$, estimated average electron density
  $n_\mathrm{e}$, and resulting minimum detectable hidden-photon mass
  $m_{\gamma_\mathrm{s}}$ as estimated from the average electron
  density.}
\label{tb:igm}
\begin{center}
\begin{ruledtabular}
\begin{tabular}{l>{\hspace{-0.2cm}}c>{\hspace{-0.1cm}}ccccc}
Baryonic & $T$ & $m/m_\mathrm{b}$ & $V/V_\mathrm{c}$  &$\rho/\rho_\mathrm{b}$   & $n_\mathrm{e}$ & $m_{\gamma_\mathrm{s}}$ \\ 
component & [K] &  & & & [cm$^{-3}$] & [eV] \\
\hline
Galaxies     & $<$10$^3$       & 0.054 & 0.002 & 27.0  & 6.7$\times$10$^{-6}$ &  6.7$\times$10$^{-14}$ \\
Warm IGM     & $<$10$^5$       & 0.350 & 0.342 & 1.02  & 2.6$\times$10$^{-7}$ &  1.3$\times$10$^{-14}$ \\
WHIM IGM     & $<$10$^6$       & 0.471 & 0.030 & 15.7  & 3.9$\times$10$^{-6}$ &  5.1$\times$10$^{-14}$ \\
Hot IGM      & $>$10$^6$       & 0.097 & 0.006 & 16.2  & 4.0$\times$10$^{-6}$ &  5.2$\times$10$^{-14}$ \\
Voids        & $\sim$$10^6$(?) & 0.028 & 0.620 & 0.05  & 1.1$\times$10$^{-8}$ &  2.7$\times$10$^{-15}$ \\
\end{tabular}
\end{ruledtabular}
\end{center}
\end{table}

In cosmic voids, strong density gradients are observed \citep[][and
  references therein]{sheth2004,pan2012}, with densities at the void
center being at least 2--3 orders of magnitude lower than at the edge
of the void. Given the galaxy density as a tracer of the gas density,
the density profile of electrons, $n_\mathrm{e}(r^\prime)$, can be
calculated from the average radial density profile obtained from the
galaxy counts (see Fig.~4--6 in \cite{pan2012}), where $r^{\prime} =
r/r_\mathrm{void}$, with $r_\mathrm{void}$ denoting the void
radius. The profile was normalized to reproduce the void average
density reported in \cite{gottloeber2003}, see Table~\ref{tb:igm}. For
such a profile, the minimum $m_{\gamma_\mathrm{s}}$ can then be
calculated by evaluating $n_\mathrm{e}(r)$ at $L_\mathrm{osc}/2$ (to
account propagation on both sides from the center of the void). The
resulting values are given in Table~\ref{tb:mmin} for several typical
values of the void size and observing frequencies.

\begin{table}
  \caption{Minimum detectable hidden-photon mass
    $m_{\gamma_\mathrm{s}}$ in eV, assuming photon
    propagation through cosmic voids.} \label{tb:mmin}
\begin{center}
\begin{tabular}{cccc}\hline\hline
Void & \multicolumn{3}{c}{Observing frequency} \\
$2r_\mathrm{void}$\,[Mpc] & 30\,MHz & 100\,MHz & 1\,GHz \\\hline
10  & 1.7$\times$10$^{-17}$ & 2.3$\times$10$^{-17}$ & 4.1$\times$10$^{-17}$ \\
30  & 1.3$\times$10$^{-17}$ & 1.7$\times$10$^{-17}$ & 3.1$\times$10$^{-17}$ \\
100 & 9.6$\times$10$^{-18}$ & 1.3$\times$10$^{-17}$ & 2.3$\times$10$^{-17}$ \\
\hline \hline
\end{tabular}
\end{center}
\end{table}

This calculation demonstrates that for the assumptions used here,
hidden photons with masses as low as $\sim\!10^{-17}$\,eV can in
principle be probed with astrophysical measurements of targets located
behind sufficiently large voids. A more detailed account of ISM/IGM
inhomogeneities on generation and propagation of the hidden photon
signal would rely on extensive numerical simulations of large scale
structures, which is beyond the scope of this paper. In the following
discussion, the generic $m_{\gamma_\mathrm{s}}$ limits obtained for
Galactic ($\gtrsim 10^{-14}$\,eV) and extragalactic ($\gtrsim
10^{-17}$\,eV) objects will be assumed, while performing calculations
for the $m_\gamma \le \sqrt{2}\,m_{\gamma_\mathrm{s}}$ regime.

It is interesting to note that the signal from hidden photons with a
sufficiently low mass, $m_{\gamma_\mathrm{s}}^\mathrm{free}$, should be
unaffected by propagation through a high-density structure with a
characteristic size $l_\mathrm{s}$, for which $l_\mathrm{s} < L_\mathrm{osc}$.
This implies $m_{\gamma_\mathrm{s}}^\mathrm{free} =
2\pi\sqrt{2}\,(\nu/l_\mathrm{s})^{1/2}$ and yields, at $\nu = 1$\,GHz,
$m_{\gamma_\mathrm{s}}^\mathrm{free}\approx 1.3 \times 10^{-16}$\,eV
for propagation through galaxies ($l_\mathrm{s}\approx 20$\,kpc).

\section{Detection of the oscillation signal}
\label{sc:detection}

For measurements in the $m_\gamma \le \sqrt{2}\,m_{\gamma_\mathrm{s}}$
regime, the two main factors limiting the sensitivity for a
hidden-photon signal are the spectral energy distribution of the
(typically, multicomponent) astrophysical signal and spectral range
covered by the resolution of astronomical instruments. As the
hidden-photon signal modulates the astrophysical signal, the latter
has to be well understood before attempting to detect the oscillation
signal in a broad-band spectrum. The astrophysical signal can be
modeled with ${\cal M}(\nu)$, such that a condition
\begin{equation}\label{eq:signal}
F^{\prime}_{\gamma_\mathrm{s}}(\nu) =
F_{\gamma_\mathrm{s}}(\nu)/{\cal M}(\nu) = C_{\cal M}
(1-P_{\gamma\rightarrow\gamma_\mathrm{s}}) + {\cal F}_\mathrm{noise}
\end{equation}
is achieved (or approached), where $C_{\cal M}$ is a constant
(expecting $C_{\cal M} \rightarrow 1$) and ${\cal F}_\mathrm{noise}$ is
the residual fractional noise due to measurement errors and systematic
uncertainties of the fit by ${\cal M}(\nu)$ (with $\sigma_\mathrm{rms}
\ll C_{\cal M}$). In the radio regime, measurement errors will be
dominated by the system noise and atmospheric/ionospheric
fluctuations, while the effect of scattering in interstellar and
intergalactic plasma would be orders of magnitude smaller and could be
safely neglected. If necessary, the condition $C_{\cal M}=1$ can be
achieved by normalizing $F^{\prime}_{\gamma_\mathrm{s}}(\nu)$ over the
observed frequency range (this measure would increase the noise and
potentially introduce a bias, but it may be necessary to facilitate
subsequent searches for a periodic signal).

The conversion probability $P_{\gamma\rightarrow\gamma_\mathrm{s}}$ is
periodic in the wavelength domain, which requires that the model
description ${\cal M}(\nu)$ must not contain harmonic terms within a
certain frequency range. This range is determined by several specific
factors, including the distance to the object and the specific value
of $m_{\gamma_\mathrm{s}}$ to be probed. This range is calculated and
discussed below.

In the following, it is assumed that the observational setup for a
broad-band spectrum measurement can be simplified and characterized by
a range $[\nu_1; \nu_2]$ of frequencies probed with a spectral
resolution $\Delta \nu$ (implying that the flux density measurements
are made at average intervals $\Delta \nu$). In general, $\Delta\nu$
may vary across the frequency range, hence it is used here only in the
sense of defining the total number of independent flux density
measurements $N_\mathrm{mes} =(\nu_2-\nu_1)/\Delta\nu$.

\subsection{Effective ranges of frequency and hidden-photon mass}
The sensitivity to detect the imprint of the amplitude $a_\chi$ with a
significance of $n_\sigma$ on an observed radio spectrum with a
Gaussian noise described by $\sigma_\mathrm{rms}$ is given by $a_\chi =
n_\sigma\,\sigma_\mathrm{rms}$ (neglecting for the moment systematic
errors resulting from an imperfect model representation, ${\cal
  M}(\nu)$, of the astrophysical signal and systematic uncertainties
of the measurement process). The highest radio frequency,
$\nu_\mathrm{h}$, useful for recovering the oscillation signal can be
estimated from $P_{\gamma\rightarrow\gamma_\mathrm{s}} =
\sigma_\mathrm{rms}$, yielding (for $n_\sigma \ge 1$, using
Eq. \ref{eq:osc_prob})
\begin{equation}
\nu_\mathrm{h} = \frac{\pi}{2} \frac{\nu_\star}{\arcsin(1/\sqrt{n_\sigma})}\,,
\end{equation}
with $\nu_\mathrm{h} = \nu_\star$ at $n_\sigma =1$. At $n_\sigma \ge
2$, $\nu_\mathrm{h} \approx (\pi/2)\,n_\sigma^{1/2} \nu_\star$ gives
an estimate of $\nu_\mathrm{h}$ to within a 10\% accuracy.

The lowest frequency, $\nu_\ell$, containing a usable response from
the oscillation is determined by the spectral spacing, $\Delta
\nu$. In this case, a conservative estimate of $\nu_\ell$ is provided
by the double of the Nyquist sampling rate, $f_\mathrm{s}$, of the
oscillation signal (this is required to take into account that
measurements are made at a fixed set of frequencies and hence no
``phase tuning'' is feasible). The requirement corresponds to
$\varphi_{\gamma_\mathrm{s}}(L,\nu_\ell-\Delta\nu/2)-\varphi_{\gamma_\mathrm{s}}(L,\nu_\ell+\Delta\nu/2)
= \pi/2$ and yields $\nu_\ell = \sqrt{\nu_\star \Delta\nu
  +(\Delta\nu/2)^2} \approx \sqrt{\nu_\star \Delta\nu}$.

\begin{figure}[t]
\begin{center}
\includegraphics[width=0.45\textwidth]{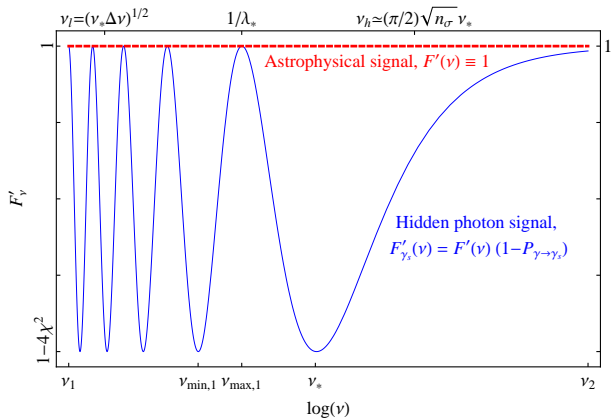}
\end{center}
\caption{Modulation of an ideally modeled astrophysical signal,
  $F^{\prime}(\nu)\equiv 1$, by the hidden photon
  oscillations. Measurements cover the $[\nu_1;\nu_2]$ frequency range
  and are made with a spectral resolution of $\Delta\nu$. Oscillations
  occur at a constant wavelength $\lambda_\star$. Detection of the
  hidden-photon signal can be made if $\nu_1 \le \nu_\mathrm{min,2}$
  and $\nu_2 \ge \nu_\star$. Effective measurements (using the full
  sensitivity of the data) can be performed if $\nu_1 \le \nu_\ell$
  and $\nu_2 \ge \nu_\mathrm{h}$. The spectral resolution must be
  better than $\nu_\star/4$.}
\label{fg:scheme}
\end{figure}

Generic properties of a possible measurement are illustrated in
Fig.~\ref{fg:scheme}, showing a modulation of an astrophysical signal
in the observed frequency range $[\nu_1;\nu_2]$, together with the
limiting frequencies $\nu_\ell$ and $\nu_\mathrm{h}$.

\begin{figure*}[t!]
\begin{center}
\includegraphics[width=0.9\textwidth]{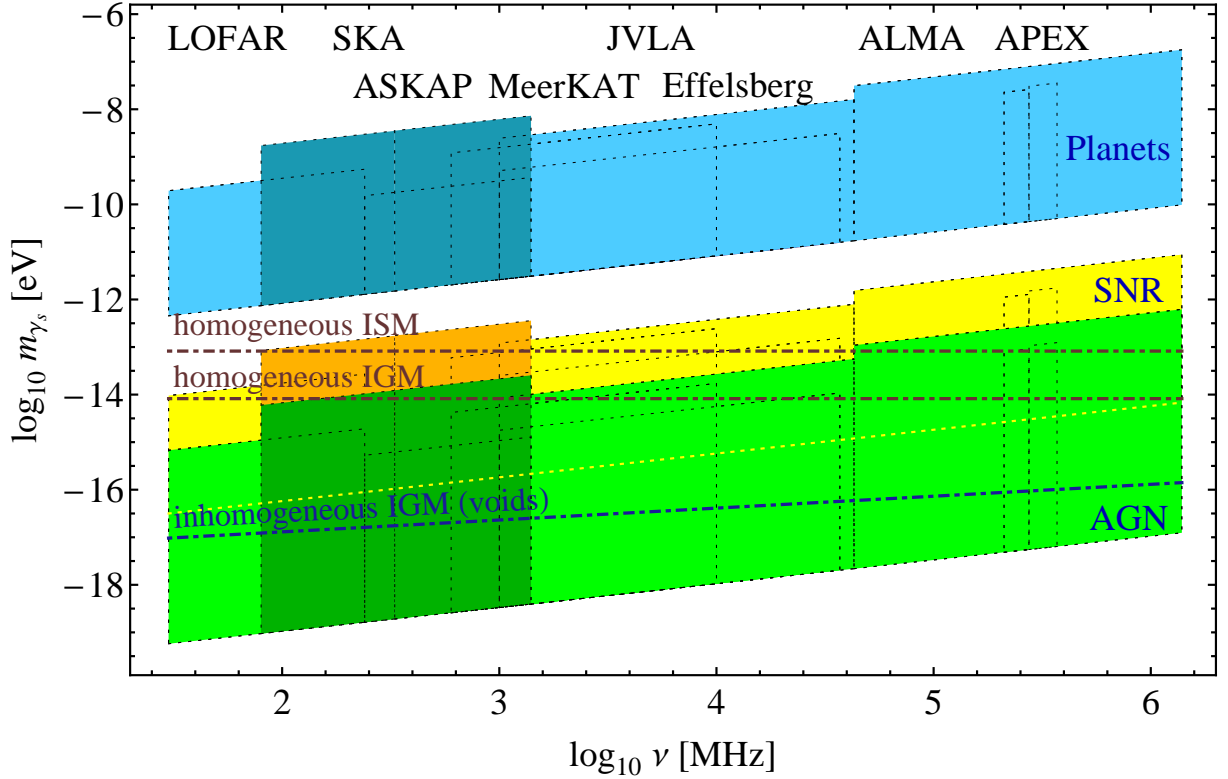}
\end{center}
\caption{Ranges of hidden-photon mass, $m_{\gamma_\mathrm{s}}$, that
  can be probed with various radio astronomical instruments at
  different observing frequencies. The mass ranges are calculated for
  measurements done with planets (blue shades), supernovae remnants
  (yellow shades) and active galactic nuclei (green shades).  For each
  individual color, darker shades mark the mass ranges accessible to
  measurements with the first and second phases of Square Kilometer
  Array (SKA). The calculations are made assuming typical instrumental
  setups and generic ranges of distances to planets (0.5--10\,au),
  supernova remnants (1--10\,kpc) and active galaxies (0.02--3\,Gpc).
  Brown dot-dashed lines mark the lower limits on detectable mass
  imposed by homogeneous ISM and IGM suppression. The dashed blue line
  illustrates the lower limit on detectable mass arising from
  propagation throuh an inhomgeneous IGM containing large-scale voids
  (for the assumed void diameter of 100\,Mpc).}
\label{fg:masses}
\end{figure*}

The characteristic frequencies translate into an accessible mass range
extending from $m^\ell_{\gamma_\mathrm{s}}$ to
$m^u_{\gamma_\mathrm{s}}$. The lower limit on the detectable
hidden-photon mass $m^\ell_{\gamma_\mathrm{s}}$ is determined by $L=
L_\mathrm{osc}(\nu_1)$ (see Section \ref{ssec:osc_coh}), corresponding
to $\nu_1$ set to the frequency of the first local maximum
$\nu_\mathrm{max,1}$. This yields
\begin{eqnarray}
m^\ell_{\gamma_\mathrm{s}} & = & 2\sqrt{2}\,\pi
\left(\frac{\nu_1}{L} \right)^{1/2} \nonumber \\
 & = & 5.77 \times 10^{-16}
\left(\frac{\nu_1}{\mathrm{MHz}}\right)^{1/2}
\left(\frac{L}{\mathrm{pc}}\right)^{-1/2}\mathrm{eV}\,.
\label{eq:mgll}
\end{eqnarray}
The largest accessible hidden-photon mass
$m^\mathrm{u}_{\gamma_\mathrm{s}}$ is obtained by requiring $\nu_\ell
= \nu_2$, which gives
\begin{equation}\label{eq:mgul}
m^\mathrm{u}_{\gamma_\mathrm{s}} = \frac{2\pi\,\nu_2}{(L\,
\Delta\nu)^{1/2}} = \frac{1}{\sqrt{2}}
\frac{\nu_2}{\nu_1} \left( \frac{\Delta \nu}{\nu_1} \right)^{-1/2}
m^\ell_{\gamma_\mathrm{s}}\,.
\end{equation}
These limits are analyzed and presented in Fig.~\ref{fg:masses} for
several existing and planned radioastronomical facilities, and for
different types of astrophysical targets. In real experiments, the
accessible mass ranges may be further limited by coherence effects and
medium propagation as discussed in Sections \ref{ssec:osc_coh} and
\ref{sc:detection}. This is illustrated in Fig.~\ref{fg:masses} by
comparing the accessible ranges of hidden-photon masses to the limits
imposed by the homogeneous ISM and IGM suppression. The impact of the
medium suppression can be alleviated at lower photon masses by free
propagation through a homogeneous medium (also illustrated in
Fig.~\ref{fg:masses}), in addition to the favorable conditions that
may exist for propagation through an inhomogeneous medium.

The sensitivity to $\chi$ varies within the mass ranges described by
Eqs.~\ref{eq:mgll} and \ref{eq:mgul}.  The lowest hidden-photon mass
$m^{\ell,\mathrm{full}}_{\gamma_\mathrm{s}}$ for which a set of
measurements made in the [$\nu_1;\nu_2$] frequency range is fully
sensitive to $\chi$ is set by the condition $\nu_\ell = \nu_1$, which
corresponds to
\begin{equation}
m^{\ell,\,\mathrm{full}}_{\gamma_\mathrm{s}} =
\frac{2\pi\,\nu_1}{(L\,\Delta\nu)^{1/2}} = \frac{1}{\sqrt{2}}
 \left( \frac{\Delta \nu}{\nu_1} \right)^{-1/2}
m^\ell_{\gamma_\mathrm{s}}\,.
\label{eq:mglr}
\end{equation}
The largest mass that can be detected at the full sensitivity is then
given by the condition $\nu_\mathrm{h} = \nu_2$, resulting in
\begin{equation}
m^{\mathrm{u},\,\mathrm{full}}_{\gamma_\mathrm{s}} = 2\sqrt{2\pi}
\left[\frac{\nu_2}{L} \arcsin\left(\frac{1}{\sqrt{n_\sigma}}\right) \right]^{1/2}\,,
\label{eq:mgur}
\end{equation}
with $m^{\mathrm{u},\,\mathrm{full}}_{\gamma_\mathrm{s}} = 2\pi
(\nu_2/L)^{1/2}$ for $n_\sigma = 1$ and
$m^{\mathrm{u},\,\mathrm{full}}_{\gamma_\mathrm{s}} \approx
2\sqrt{2\pi} (\nu_2/L)^{1/2} n_\sigma^{-1/4}$ for $n_\sigma \ge 2$.
With this approximation,
$m^{\mathrm{u},\,\mathrm{full}}_{\gamma_\mathrm{s}}$ can be expressed
through $m^\ell_{\gamma_\mathrm{s}}$
\begin{equation}
m^{\mathrm{u},\,\mathrm{full}}_{\gamma_\mathrm{s}} \approx
\frac{1}{\sqrt{\pi}\,n_\sigma^{1/4}}\left(\frac{\nu_2}{\nu_1}\right)^{1/2}
m^\ell_{\gamma_\mathrm{s}}\,.
\label{eq:mgurapprox}
\end{equation}
The mass limits described by Eqs.~\ref{eq:mgll}-\ref{eq:mgur} can be
used to estimate the frequency dependence (or, conversely, the
hidden-photon mass dependence) of the sensitivity provided by a given
set of measurements for detecting the hidden-photon signal.

Eqs.~\ref{eq:mglr} and \ref{eq:mgur}, in particular, can be used for
experiment optimization by requiring that
$m^{\mathrm{u},\,\mathrm{full}}_{\gamma_\mathrm{s}}>
m^{\ell,\,\mathrm{full}}_{\gamma_\mathrm{s}}$, which corresponds to the inequality
\[
\sqrt{\frac{2}{\pi}}\, \frac{1}{n_\sigma^{1/4}} \left(
\frac{\nu_2}{\nu_1} \right)^{1/2} \left( \frac{\Delta \nu}{\nu_1}
\right)^{1/2} > 1
\]
and connects the main parameters of the observational setup. For the
goal of extending hidden photon studies to progressively lower
hidden-photon masses, the most efficient strategy would therefore be
to reduce $\nu_1$. Improving the frequency spacing can result in
producing a progressively larger number of datapoints that could not
be used for probing higher hidden-photon masses. Hence, the overall
range of full sensitivity of a given experimental setup would be
reduced in this case.

\subsection{Sensitivity for the mixing angle $\chi$}
\label{sc:limits-chi}

The amplitude term $a_\chi$ of $P_{\gamma \rightarrow
  \gamma_\mathrm{s}}$ implies that an $n_\sigma$ bound on $\chi
\approx \sqrt{n_\sigma \sigma_\mathrm{rms}}/2$ can be obtained from
multi-frequency flux density measurements. For $N$ individual flux
density measurements with fractional errors $\sigma_i$,
$\sigma_\mathrm{rms} = \left(\sum_{i=1}^{N} \sigma_i^2\right)^{1/2}/N
\approx \sigma/N^{1/2}$ (if $\sigma_i \approx \sigma$ for all
$i=1,\dots,N$).  One can assume, as an example, that the astrophysical
signal from a target object can be described by a simple power-law
spectrum $F(\nu) = F_\mathrm{r} (\nu/\nu_\mathrm{r})^\alpha$ and that
the frequency dependence of the errors on flux density measurements
changes can also be described by a power-law dependence $\sigma(\nu) =
\sigma_\mathrm{r} (\nu/\nu_\mathrm{r})^\beta$ (here, $F_\mathrm{r}$
and $\sigma_\mathrm{r}$ refer to a flux density and its associated
error, measured at a reference frequency $\nu_\mathrm{r}$ chosen
inside the relevant frequency range).

\begin{figure*}[t]
\begin{center}
\includegraphics[width=0.48\textwidth,clip=true]{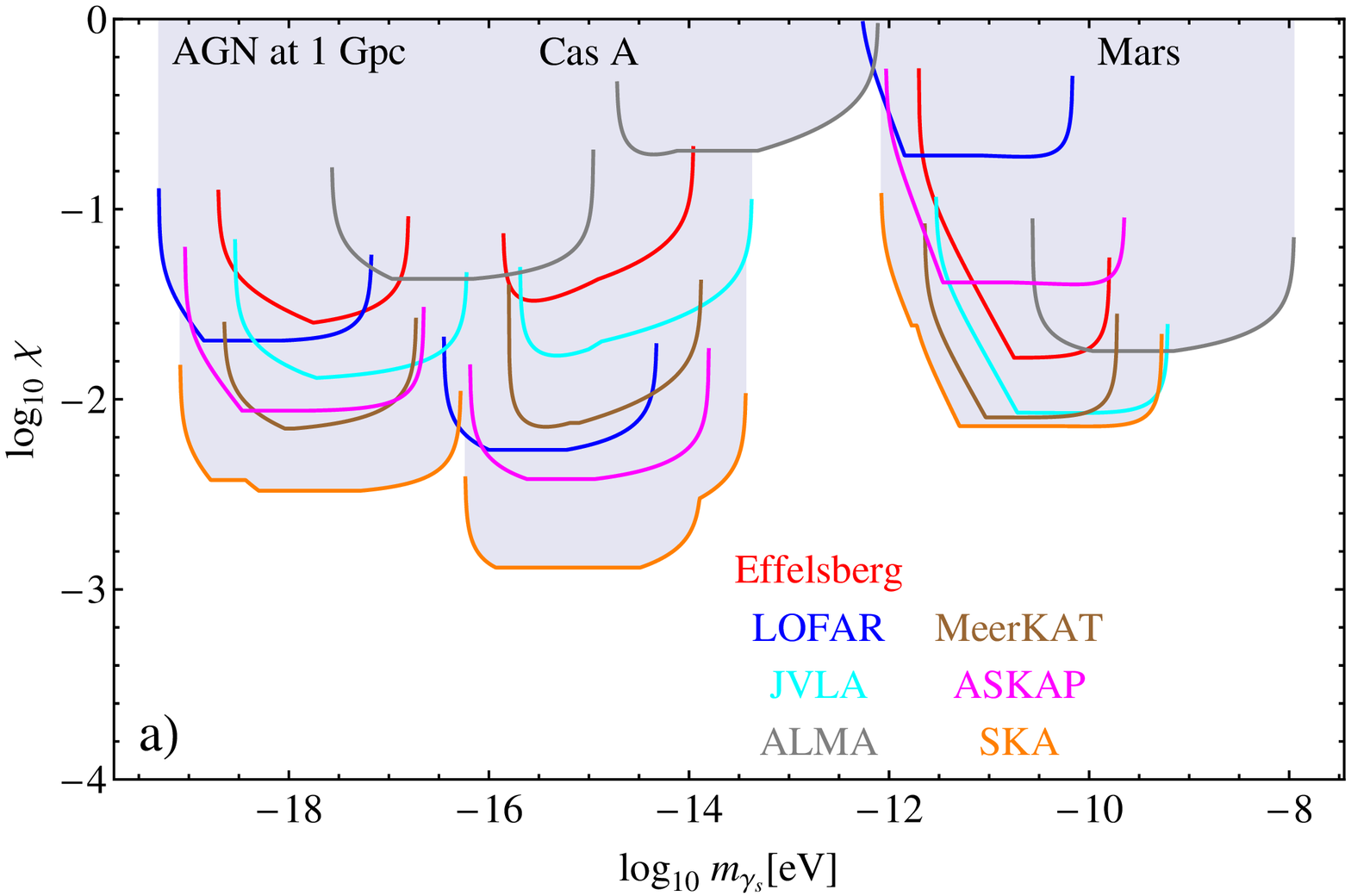}
\includegraphics[width=0.48\textwidth,clip=true]{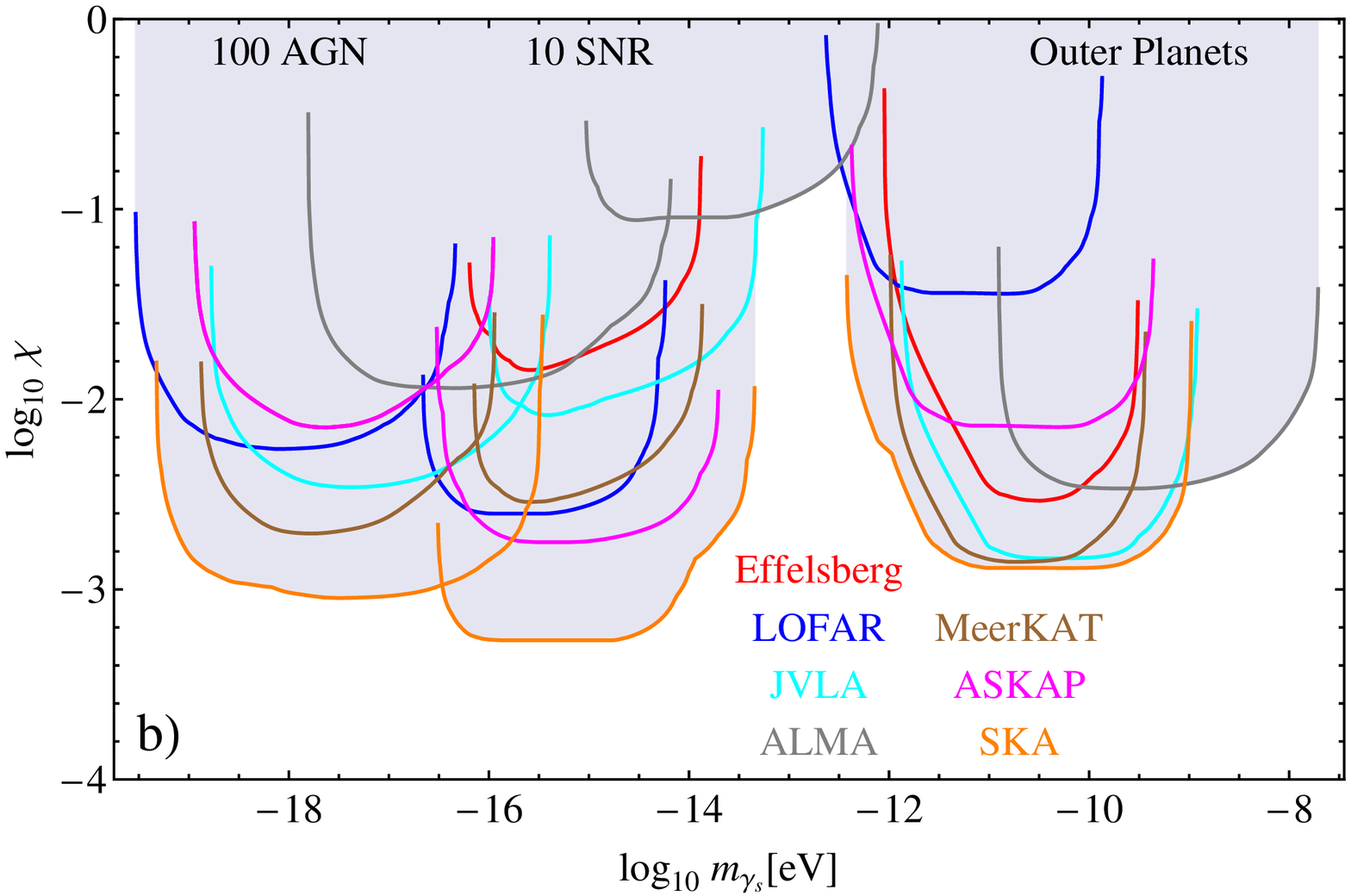}
\end{center}
\caption{Expected limits on $\chi$ that can be obtained with different
  radio astronomical facilities and different astrophysical
  objects under the assumption that the
  measurements are not affected by the plasma propagation effects (vacuum oscillations). The limits are calculated assuming generic instrumental
  parameters and: (a) single target sources at distances of 1\,Gpc,
  3.4\,kpc \cite{fesen2006} and 2\,au for the AGN, Cas~A, and Mars,
  respectively, (b) stacked data limits obtained from simulated
  populations of 100 AGN ($L=$0.02--3\,Gpc, $F=$1--30\,Jy), 10 SNR
  ($L=$1--10\,kpc, $F=$10--100\,Jy) and 100 measurements made for the
  outer planets ($L=$0.5--10\,au, $F=$10--200\,Jy), with $\gamma=-2$
  used for all three populations.}
  \label{fg:lims}
\end{figure*}

With these assumptions, measurements over the entire relevant
frequency range $[\nu_{\ell^\prime};\nu_\mathrm{h^{\prime}}]$
can be described by a characteristic signal-to-noise ratio
\begin{equation}
\hat{S} = \frac{1}{\nu_\mathrm{h^{\prime}}-\nu_{\ell^\prime}} \int\limits_{\nu_{\ell^\prime}}^{\nu_\mathrm{h^\prime}}
\frac{F(\nu)}{\sigma(\nu)} \mathrm{d}\nu = 
\frac{F_\mathrm{r}}{\sigma_\mathrm{r}\,\nu_\mathrm{r}^{\xi}}\frac{\nu_\mathrm{h^{\prime}}^{\xi+1}-\nu_{\ell^\prime}^{\xi+1}}{(\xi+1)(\nu_\mathrm{h^{\prime}}-\nu_{\ell^\prime})}\,,
\end{equation}
where $\xi = \alpha-\beta$ and the integration limits are given by
$\nu_{\ell^\prime} = \mathrm{max}(\nu_1,\nu_\ell)$ and
$\nu_\mathrm{h^{\prime}} = \mathrm{min}(\nu_2,\nu_\mathrm{h})$. At a
frequency spacing $\Delta\nu$, the number of measurements contributing
to the detection is $N_\mathrm{mes} =
(\nu_\mathrm{h^{\prime}}-\nu_{\ell^\prime})/\Delta\nu$, hence the
effective cumulative signal-to-noise ratio of the data set is
$\tilde{S} = N_\mathrm{mes}^{1/2}\,\hat{S}$. Recalling that, after
accounting for the astrophysical signal, $\tilde{S} \approx
1/\sigma_\mathrm{rms}$ gives an estimate of the lowest achievable
bound on the hidden-photon coupling
\begin{equation}
\chi_\mathrm{low} = \left(\frac{\sigma_\mathrm{r}\,\nu_\mathrm{r}^{\xi}}{4 F_\mathrm{r}}
\frac{(\xi+1)(\nu_\mathrm{h^{\prime}}-\nu_{\ell^\prime})^{1/2}(\Delta\nu)^{1/2}}{\nu_\mathrm{h^{\prime}}^{\xi+1}-\nu_{\ell^\prime}^{\xi+1}}
\right)^{1/2}\,.
\label{eq:chilow1}
\end{equation}
The bound $\chi_\mathrm{low}$ remains constant for hidden-photon
masses $m^{\ell,\,\mathrm{full}}_{\gamma_\mathrm{s}} \le
m_{\gamma_\mathrm{s}} \le
m^{\mathrm{u},\,\mathrm{full}}_{\gamma_\mathrm{s}}$ and decreases
rapidly outside this mass range as a progressively larger fraction of
the measured data points are rendered outside the useful ranges of
frequencies.

Figure~\ref{fg:lims}a presents limits $\chi_\mathrm{low}$ calculated
for several existing and planned radio astronomical facilities and for
measurements made with Mars (with $F_\mathrm{r}=30$\,Jy and
$\alpha=0.7$ at $\nu_\mathrm{r}=86$\,GHz; \cite{weiland2011}), the
supernova remnant Cassiopeia~A ($F_\mathrm{r}=3000$\,Jy, $\sigma_
\mathrm{rms}=100$\,Jy,
$\nu_\mathrm{r}=1$\,GHz, $\alpha=-0.8$; \cite{baars1972}), and a
fiducial compact AGN ($F_\mathrm{r}=10$\,Jy, $\alpha=-0.1$,
$\nu_\mathrm{r}=5$\,GHz) at a distance of 1\,Gpc. These limits are
obtained for the vacuum oscillation regime, without taking into
account the potential medium suppresion at lower hidden photon
masses. For each of the instruments included in the plot, conservative
assumptions for generic technical parameters (summarized in
Appendix~B) have been made, hence the actual limits could be further
improved by optimizing observations with a given telescope, for
instance by increasing the spectral resolution of the measurements or
applying accurate in-band (bandpass) calibration \cite{winkel2012}.
The figure clearly reflects the effect of improvements of
$\sigma_\mathrm{rms}$ (factor of $\sim 10$) and $\Delta\nu/\nu$
(factor of $\sim 100$) that will be provided by the JVLA and SKA
precursors, as well as the extension (factor of $\sim 10$) to lower
frequencies provided by LOFAR.

\subsection{Source stacking}
\label{sc:limits-stacks}

Since the frequency behavior of the hidden-photon signal is determined
solely by the hidden-photon mass and the distance to the target
object, signals from any number of objects with known distances can be
stacked together to improve the resulting bound on the coupling
constant. For the stacking, a suitable reference distance $L^{\prime}$
can be chosen, yielding for each object located at a distance $L$ a
modification of observed frequency $\nu^{\prime} =
\nu\,(L^{\prime}/L)$.

If all of the stacked sources have the same spectra and distances,
stacking of $N_\mathrm{obj}$ objects will lead to a
$N_\mathrm{obj}^{1/4}$ improvement of $\chi_\mathrm{low}$.  Let the
stacked objects be drawn from a population with a uniform spatial
density over distances $[L_\mathrm{min};L_\mathrm{max}]$, similar
spectral indices, and an observed source count, $N(F) =
n(F_\mathrm{r}) (F/F_\mathrm{r})^{\gamma}$ over a range of flux
densities $[F_\mathrm{min};F_\mathrm{max}]$.

In this case, stacking of $N_\mathrm{obj}$ spectra (each obtained with
the same observational apparatus as described above) would yield
a bound
\begin{equation}
\chi_\mathrm{stack} = \chi_\mathrm{low} N(\hat{F})^{-1/4}\,,
\end{equation}
where $\hat{F}$ is estimated at a frequency $\hat{\nu}$ at which the
signal-to-noise ratio $\hat{S}$ is achieved. Calculation of $\hat{S}$
may now involve different integration limits, as the frequencies
$\nu_\ell$ and $\nu_\mathrm{h}$ must be calculated for a
characteristic distance $\hat{L}$. This distance is given by
$[(L_\mathrm{max}^2 + L_\mathrm{min}^2)/2]^{1/2}$ for planets and Galactic
objects and by $[(L_\mathrm{max}^3 + L_\mathrm{min}^3)/2]^{1/3}$ for
extragalactic objects.

For the source and observation properties specified by $F(\nu)$ and
$\sigma(\nu)$ (see Section~\ref{sc:limits-chi}), $\hat{\nu}
  = \nu_\mathrm{r} (\hat{S}
  \sigma_\mathrm{r}/F_\mathrm{r})^{1/\xi}$, giving
$\hat{F} =
  F_\mathrm{r}^{1-\alpha/\xi} (\hat{S}
  \sigma_\mathrm{r})^{\alpha/\xi}$. Consequently,
$N(\hat{F}) = n(F_\mathrm{r}) (\hat{F}/F_\mathrm{r})^{\gamma}$,
where $n(F_\mathrm{r}) = N_\mathrm{obj} F_\mathrm{r}^{\gamma}
(1+\gamma) \left(F_\mathrm{max}^{1+\gamma} -
F_\mathrm{min}^{1+\gamma}\right)^{-1}$.

\begin{figure*}[t!]
\begin{center}
\includegraphics[width=0.9\textwidth]{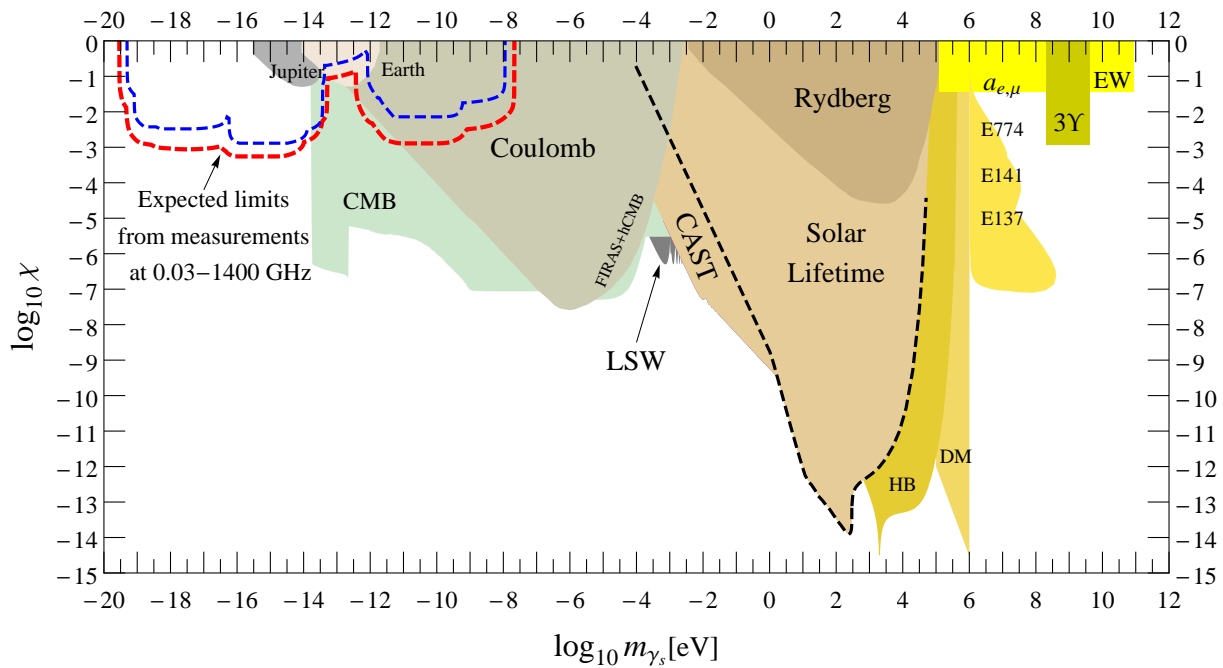}
\end{center}
\caption{Compound limits on $\chi$ expected to be achievable from
  observations at 0.03--1400\,GHz compared with the limits obtained
  presently with other facilities and experiments \citep[][and
  references therein]{jaeckel2010}. The limits from single object
  (blue) and multiple object stacking (red) are shown. Radio
  observations (particularly at frequencies below 40\,GHz) will
  provide a unique probe for the hidden photon with masses below
  $10^{-14}$\,eV and extending down to $\sim 10^{-17}\,\mathrm{eV}$
  where the measurements are likely to be limited by the medium
  suppression of the hidden-photon signal in the IGM plasma as
  described in Sections~\ref{sc:refr}--\ref{sc:inhom}.}
\label{fg:all_limits}
\end{figure*}

Potential improvements of source stacking are illustrated in
Fig.~\ref{fg:lims}b which shows the limits on $\chi$ that can be
achieved by stacking together measurements made for 100 AGN, 10 SNR,
and 100 measurements obtained for different outer planets of the Solar
System. Improvements in both $\chi$ and the range of accessible
hidden-photon mass are visible, compared to the single object limits
in Fig~\ref{fg:lims}a. The predictions for best cumulative limits from
radio measurements in the entire 30 MHz to 1400 GHz range are compared
in Fig.~\ref{fg:all_limits} to the bounds derived from other
experiments and observations.  Fig.~\ref{fg:all_limits}
  demonstrates that radio measurements would be unique for detecting
  hidden photons with $m_{\gamma_\mathrm{s}}\lesssim 3 \times
  10^{-16}$\,eV and would substantially improve the existing limits
  for hidden photon masses below $2\times 10^{-14}$\,eV which is the
  lowest mass for which the resonant conversion can be assessed from
  the FIRAS CMB measurements \cite{mirizzi2009a}. The limits based on
  the FIRAS data can in principle be extended to photon masses as low
  as $\approx 7\times 10^{-17}$\,eV, if non-resonant conversion is
  considered. However, even in this case, the radio measurements would
  provide stronger constrains on $\chi$ for $m_{\gamma_\mathrm{s}}\lesssim
  10^{-15}$\,eV.

\begin{figure*}[t!]
\begin{center}
\includegraphics[width=0.9\textwidth]{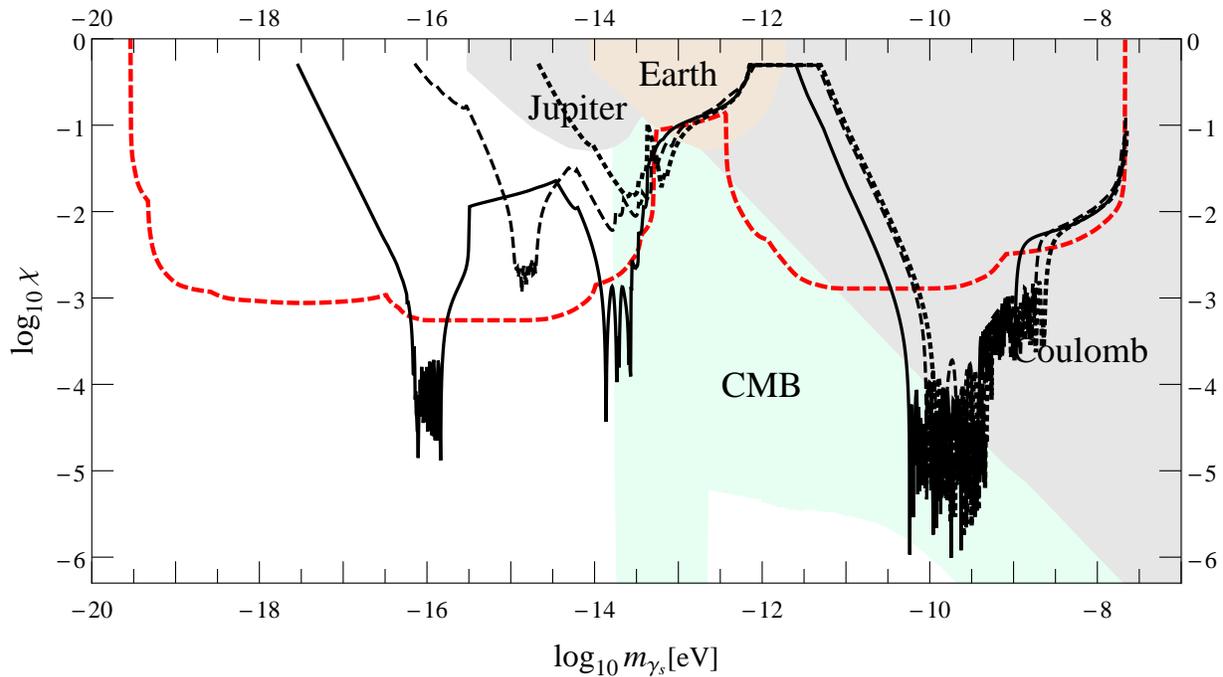}
\end{center}
\caption{Modifications of the combined limits on $\chi$ due to
  propagation through refractive media. The dashed red line shows the
  limits predicted for object stacking observations under the vacuum
  condition (the same as the dashed red line in
  Fig.~\ref{fg:all_limits}).  The best (solid black), average (dashed
  black) and worst (dotted black) propagation scenarios illustrate
  potential effects of the ISM and IGM plasma for constraining the
  hidden-photon signal. For the best case scenario, the effect of
  resonant enhancement is visible. The medium suppression affects
  strongly the limits on $\chi$ for hidden-photon masses below $\sim
  10^{-17}$\,eV.}
\label{fg:med_limits}
\end{figure*}

The potential effect of propagation through refractive media on the
hidden-photon limits is illustrated in Fig.~\ref{fg:med_limits} for
three different cases describing the minimum values of the plasma
density, $n_\mathrm{e,min}(l)$, as a function of distance, $l$, along
the propagation path. The three scenarios for $n_\mathrm{e,min}(l)$ are
adopted here solely to assess the potential range of possible outcomes of
the propagation of the hidden-photon signal through refractive media. 

In all three scenarios,
$n_\mathrm{e,min}(\mathrm{1\,kpc})=10^{-5}\,\mathrm{cm}^{-3}$ is
adopted, implying effectively that the LOS paths for Galactic objects
cross at least one of the Galactic ``mini void'' regions.  The worst
case scenario assumes that the photon signal from extragalactic
objects propagates mostly through Galactic ISM, {\em i.e.}, not
crossing large patches of IGM. In this case, $n_\mathrm{e,min}$
decreases from $10^{-5}\,\mathrm{cm}^{-3}$ at $l=1$\,kpc to
$10^{-7}\,\mathrm{cm}^{-3}$ at $l=3$\,Gpc implying that, for more
distant objects, the LOS path has a progressively larger probability
to cross lower density IGM regions. The average case corresponds to
propagation through typical IGM resulting in a $n_\mathrm{e,min} \sim
10^{-5}$--$10^{-8}\,\mathrm{cm}^{-3}$ range of densities. The best
case allows for propagation through cosmic voids, with the
corresponding $n_\mathrm{e,min} \sim
10^{-5}$--$10^{-10}\,\mathrm{cm}^{-3}$.  For the planets, the
densities of 50, 30, and 10\,cm$^{-3}$ (measured at 1\,au and scaling
as $\propto (l/\mathrm{1\,au})^{-2}$) have been adopted for the
respective scenarios, based on measurements from \cite{tappin1986}.

The effects of resonant enhancement and medium suppression are clearly
visible in the modified limits shown in
Fig.~\ref{fg:med_limits}. Assuming that the best case scenarios would
apply for the majority of the lines of sight (since it is likely that
a photon beam from a distant galaxy crosses one or more rarefied IGM
or void regions), it is reasonable to conclude that the medium
suppression would reduce the detectable $\chi$ to $\ge 0.01$ only for
estimates made for $m_{\gamma_\mathrm{s}} \lesssim
10^{-17}$\,eV. Above this mass, the propagation effects should not
pose severe problems for constraining $\chi$ (and they indeed may even
play a constructive role at least for some fraction of the photon mass
range).

\subsection{Detection of periodic modulations}
\label{sc:period}

The modulations induced by the hidden-photon signal on the broad-band
spectrum of an astrophysical object can be best detected in the
wavelength domain, where the modulation is sinusoidal, with the period
given by $\lambda_\star$. A fast Fourier transform (FFT) can be
applied for searches in the data with dense and uniform coverage of
wavelength space. For sparsely sampled data, epoch folding
\cite{leahy1983a} or generic uniformity tests such as the Rayleigh
test \cite{brazier1994} or $Z^2_m$ test \cite{buccheri1983} can be
applied.

These searches employ Eq.~\ref{eq:signal} with the $C_{\cal M} =
1$ normalization of the residual flux density, which yields a
functional form
\begin{equation}
f_\lambda = f_\star [1+
a_\star \sin(\omega_\star \lambda + \phi_0)]
\end{equation} 
of the periodic signal to be searched for. Assuming that the residual
errors after the normalization are $\delta_\lambda =
\sigma_\mathrm{rms}/C_{\cal M}$, the parameters of the functional form
are related to the properties of the hidden-photon signal as follows:
$f_\star = 1 - a_\chi/2 + \delta_\lambda$, $a_\star = a_\chi/(2
f_\star)$, $\omega_\star = \pi \nu_\star$, $\phi_0 = \pi/2$.
Appendix~\ref{app:periodic} provides specific details of application
of three different methods (FFT, epoch folding and generic
uniformity tests) to searches for periodic signals due to hidden
photon oscillations.

\section{Discussion} \label{sc:disc}

The analysis described above demonstrates principal feasibility of
searching for a hidden-photon signal in broad-band radio spectra of
cosmic radio sources, with Galactic supernova remnants and radio-loud
AGN presenting the best opportunity for extending the measurements
below the hidden-photon mass of $10^{-14}$\,eV, where essentially no
measurements have been previously made.

The mass ranges and kinetic mixing limits accessible for these
potential searches are determined by several factors, including the
oscillation and coherence lengths of the hidden-photon signal, the
instrumental bandwidth and resolution, as well as the plasma density
changes along the line-of-sight path to the source of the photons.
The combined effect of the signal coherence and oscillation length is
expected to limit the searches to hidden-photon masses $\gtrsim
10^{-19}$\,eV, while the limits imposed by the propagation through the
cosmic plasma may increase this limit up to $\sim 10^{-17}$\,eV.

Bandwidth and spectral resolution of the existing and planned radio
astronomical facilities can support searches for hidden photons with
masses well below $10^{-20}$\,eV. The limiting instrumental aspect is
the accuracy of amplitude calibration of radio receiver, which may
limit plausible constraints on kinetic mixing to $\chi \approx
10^{-3}$. This problem may be alleviated by the advent of ultra
broad-band receivers supporting in-band measurements across bandwidths
in excess of 1 GHz. For this type of measurements, accurate bandpass
calibration could deliver in-band amplitude accuracy of $\ge 0.01$\%,
thus potentially further lovering the limits on the kinetic mixing by at
least an order of magnitude. 

For hidden-photon searches based on such in-band measurements, a
combination of targets located at significantly different distances
can be employed, profiting from the distance scaling of the
hidden-photon signal. This ensures that any oscillatory pattern
associated with a specific photon mass will be detectable only in one of
the two measurements. This is realized for two objects at a distance
ratio of $L_2/L_1 (L_2>L_1) \ge (\pi^2/4)(n_\sigma/\delta_\nu)$, where
$\delta_\nu$ is the fractional spectral resolution of the
measurements. With such arrangements, the upper limit for an amplitude
of a periodic oscillation in the bandpass obtained by dividing one of the two measured signals by the other would enable
constraining $\chi$ for the ranges of photon mass probed with either of
the two targets. 

The effect of propagation on the sensitivity introduces a dependence
on the assumed electron density $n_\mathrm{e}$ of the intervening
medium and therefore on the line of sight. For Galactic sources,
estimates of $n_\mathrm{e}$ at a specific line of sight can be
obtained from pulsar dispersion measurements
\cite{taylor1993,cordes2002}. Optical hydrogen absorption lines can be
used for assessing the line-of-sight structure of the IGM
\cite{caucci2008}, which can be applied for measurements in individual
extragalactic targets. The situation may improve substantially after
the large-scale H{\sc I} surveys planned at the LOFAR, MeerKAT, ASKAP
and the SKA \cite{abdalla2005,duffy2008,rottgering2011,duffy2012}
would deliver a very detailed picture of the IGM up to very high
redshifts.

In the absence of information about the plasma properties on
individual lines of sight, a potential remedy for the object stacking
would be to make generic assumptions on the minimum plasma density as
a function of distances to individual targets (similarly to the
approach employed in Section~\ref{sc:limits-stacks}). Despite being
inherently imprecise, this approach should still enable improving the
constraints obtained on $\chi$ from object stacking.

Data from the next generation of large-scale continuum and spectral
line surveys at radio wavelengths will provide sufficiently accurate
information about the broad-band continuum, line-of-sight distribution
of the IGM, and distances to many thousands of extragalactic radio
sources. This will make object stacking the tool of choice for the
hidden-photon searches in the radio regime and will certainly lead to
strong improvements of the limits obtained from radio data on the
kinetic mixing of hidden photons in the $10^{-14}$--$10^{-17}$\,eV
mass range.

\begin{acknowledgments}
  We kindly acknowledge helpful discussions with Alessandro Mirizzi,
  Andreas Ringwald, and G\"unter Sigl. We thank the referee for useful
  suggestions for improving the manuscript. APL acknowledges support
  from the Collaborative Research Center (Sonderforschungsbereich) SFB
  676 ``Particles, Strings, and the Early Universe'' funded by the
  German Research Society (Deutsche Forschungsgemeinschaft, DFG).
\end{acknowledgments}

\appendix 
\section{Analysis techniques for periodic signals} \label{app:periodic}
\subsection{FFT searches}

An FFT search can be employed effectively for recovering the hidden
photon signal, if measurements are made in a frequency range
$[\nu_1;\nu_2]$, corresponding to a wavelength range $\Lambda =
\lambda_1-\lambda_2$. The measurements are assumed to be sampled
densely enough to be binned into $n_\mathrm{b}=2^m$ bins, with each
bin described by the bin flux density $f_i$ and its associated error
$\delta_i$. This situation can be realized, for instance, for
observations with high spectral resolution (e.g., with LOFAR, JVLA, or
ALMA) applied to search for hidden photons with
$m_{\gamma_\mathrm{s}}> 2\pi \sqrt{\nu_2/L}$, for which $\nu_\star >
\nu_2$. The resulting $2m-1$ Fourier
coefficients are given by
\begin{equation}
|a_i|^2 = \sum_{j=1}^{n_\mathrm{b}} f_j \exp(i\omega_i \lambda_j)\,,
\end{equation}
where $\lambda_j =(j-1/2) \Lambda/n_\mathrm{b}$ is the arithmetic mean
wavelength of the $j^\mathrm{th}$ bin and $f_j$ is the respective
bin-averaged flux density. The corresponding power spectrum $w_i$ is
given by
\begin{equation}
w_i = 2 |a_i|^2 /F_\mathrm{tot}\,,
\end{equation} where $F_\mathrm{tot} = \sum_{j=1}^{n_\mathrm{b}} f_j$
is the total flux density in the bins.
The average power spectrum contribution from the measurement noise is
accounted for by the term 
\begin{equation}
\tilde{\delta} = \frac{1}{n_\mathrm{b}} \sum_{i=1}^{n_\mathrm{b}} \delta_i^2.
\end{equation} 
Adding the noise contribution and taking
into account frequency dependence of $|a_i|^2$ \cite{leahy1983a}, the
average power in the $i^\mathrm{th}$ bin can be written as
\begin{equation}
\langle w_i \rangle = \frac{\xi_\omega F_\mathrm{tot}}{2} \left( \frac{a_\star \sin\varphi_i}{\varphi_i}\right)^2 + \tilde{\delta} \,,
\end{equation}
where $\varphi_i = \pi i/(2n_\mathrm{b})$ and $\xi_\omega = 0.773$
\cite{leahy1983a} is a correction factor taking into account the
finite bin width.  Reaching, in a given bin, a desired confidence
level, $c$ (in percent), of detection of the oscillations, implies
$\langle w_i \rangle \ge w_0$,
where $w_0$ is derived from the $\chi^2$ probability distribution,
$p_2(\chi^2)$, with two degrees of freedom, requiring that
\begin{equation}
1- c/100 = (\Lambda/\lambda_\star)
\int_{w_0}^{\infty} p_2(\chi^2) \mathrm{d}\chi^2\,.
\end{equation} 
If a signal is
detected, the amplitude $a_\star$ (and consequently, the kinetic
mixing parameter $a_\chi$) can be obtained by requiring that 
\begin{equation}
c/100 =
\int_{w_\mathrm{a}}^{\infty} p_2(\chi^2)\mathrm{d}\chi^2\,,
\end{equation} 
with
$w_\mathrm{a} = w_0 - \langle w_i\rangle + \tilde{\delta}/2$.

\subsection{Epoch folding}

If the data are not homogeneously sampled across the measured
wavelength domain $\Lambda$ but still can be divided into phase bins
of size $\le \Delta\nu/\nu_2$ (thus giving $n_\phi \ge
2\nu_2(\nu_2-\nu_1)/(\nu_1\Delta\nu)$ phase bins), epoch folding
\citep[{\em cf.},][]{leahy1983a,leahy1983b} can be effectively
performed in order to search for all hidden-photon masses in the
$[m_{\gamma_\mathrm{s}}^\ell;m_{\gamma_\mathrm{s}}^{\mathrm{u}}]$
range. After epoch folding with a given trial wavelength
$\lambda^{\prime}$, each bin is characterized by the bin flux density
$f_i$ and its associated error $\delta_i$.  The significance of the
signal can be assessed using the statistics $s = \sum_{i=1}^{n_\phi -1}
(f_i - \tilde{f})^2/\delta_i^2$, where $\tilde{f} =
F_\mathrm{tot}/n_\phi$. For the same assumptions as used above for the
FFT searches, the mean value of the statistics is given by
\begin{equation}
\langle s \rangle = \frac{\xi_\phi F_\mathrm{tot}}
{2} \left( 
\frac{a_\star \sin\varphi}{\varphi}
\right)^2 + \frac{\tilde{\delta}}{2}\,,
\end{equation}
with $\varphi = \pi/n_\phi$ and $\xi_\phi = 0.935$
\cite{leahy1983a}. The sensitivity to oscillations is established at a
confidence level $c$ by requiring $\langle s \rangle \ge s_0$ where
$s_0$ is obtained from the condition
\begin{equation}
1 - c/100 = (\Lambda/\lambda^{\prime}) \int_{s_0}^{\infty} p_{n_{\phi} -1}(\chi^2)\mathrm{d}\chi^2\,.
\end{equation}
Similarly, the amplitude of the detected signal can be estimated from 
\begin{equation}
c/100 = \int_{s_\mathrm{a}}^{\infty} p_{n_{\phi}-1}(\chi^2)\mathrm{d}\chi^2\,,
\end{equation} 
where $s_\mathrm{a} = s_0 - \langle s\rangle + \tilde{\delta}/2$.

\subsection{Generic uniformity tests}

For poorly and unevenly sampled data, generic uniformity tests such as
the Rayleigh test \cite{leahy1983b,brazier1994,mardia2000}, the
$Z_m^2$ test \cite{buccheri1983} or the $H$-test \cite{dejager1989}
can be applied, relieving also the requirement to bin the data before
searching for a periodic signal.

Similar to the case of epoch folding, the data consisting of $n$ flux
density measurements have to be first normalized by a factor of
$f_\mathrm{min} = \min(f_i)$ and folded with a trial wavelength
$\lambda^{\prime}$, yielding a set of amplitudes $f^{\prime}_i =
f_i/f_\mathrm{min}$ and phases $\phi_i$, with $i=1,..,n$.  Calculation
of the respective trigonometric moments is done using the terms
$f^{\prime}_i \cos \phi_i$ and $f^{\prime}_i \sin \phi_i$, so that the
resulting Rayleigh power becomes
\begin{equation}
n R^2  = \frac{1}{n} \left[ \left( \sum_{i=1}^{n} f^{\prime}_i \cos \phi_i \right)^2 + \left( \sum_{i=1}^{n} f^{\prime}_i \sin \phi_i \right)^2 \right]\,.
\end{equation}
Similar modification should be done to the trigonometric moments
entering the $Z^2_m$ statistics
\begin{equation}
Z_m^2 = \frac{2}{n}\sum_{j=1}^{m} \left\{ \left[ \sum_{i=1}^{n} f^{\prime}_i \cos(j\,\phi_i) \right]^2 + \left[ \sum_{i=1}^{n} f^{\prime}_i \sin(j\,\phi_i) \right]^2 \right\}\,.
\end{equation}
The resulting calculated powers $2n R^2$ and $Z_m^2$ should be tested
against $\chi^2$ distributions $p_2(\chi^2)$ and $p_{2m}(\chi^2)$,
respectively. Adopting the same approach as for the epoch folding,
these values will yield confidence limits for detecting a periodic
signal with the wavelength $\lambda^{\prime}$.

\section{Basic technical characteristics of simulated radio observations}
\label{sc:app-tels}

\begin{table}[t!]
  \caption{Generic technical parameters of radio telescopes used in
    the calculations of sensitivity for $\chi$ in the radio
    regime.} \label{tb:tels}
\begin{center}
\begin{tabular}{cccrlr}\hline\hline
Telescope & \multicolumn{1}{c}{$\nu_1$} & \multicolumn{1}{c}{$\nu_2$} & \multicolumn{1}{c}{$N_\mathrm{ch}$} & \multicolumn{1}{c}{$\sigma_\mathrm{r}$} & \multicolumn{1}{c}{ $\beta$} \\
           & \multicolumn{1}{c}{[GHz] } & \multicolumn{1}{c}{[GHz]  } &                  & \multicolumn{1}{c}{[Jy]} & \\ \hline
LOFAR      & 0.03   & 0.24    & 1000 &  0.3  & -0.25 \\
SKA$_1$    & 0.08   & 0.33    & 5000 &  0.04 & -0.1 \\
ASKAP      & 0.1    & 1.4     & 2000 &  0.1  & -0.25 \\
SKA$_2$    & 0.3    & 3.0     & 5000 &  0.04 & 0.1   \\
MeerKAT    & 0.6    & 10      & 200  &  0.03 & 0.25 \\
Effelsberg & 0.3    & 37      & 40   &  0.1  & 0.25 \\
JVLA       & 0.3    & 43      & 500  &  0.1  & 0.25 \\
APEX       & 170    & 410     & 4000 &  10   & 0.25 \\
ALMA       & 86     & 1389    & 5000 &  1    & 0.25 \\
\hline \hline
\end{tabular}
\end{center}
\end{table}

Table~\ref{tb:tels} describes general technical parameters adopted for
simulating observations with the radio telescopes used for the
predictions of the hidden-photon mass ranges and the limits on $\chi$
from measurements in the radio regime at frequencies of
0.03--1400\,GHz. The parameters presented in the table are: the lowest
$\nu_1$ and highest $\nu_2$ observing frequencies, the number of
measurements, $N_\mathrm{ch}$, available within the observing range,
the r.m.s. noise, $\sigma_\mathrm{r}$ of a single measurement at the
reference frequency $\nu_\mathrm{r}$, and the power index $\beta$
describing the frequency dependence of the r.m.s. noise. The parameter
values given in Table~\ref{tb:tels} provide only basic benchmark
description of technical capabilities of the individual
telescopes. Conservative estimates have been adopted for the number of
spectral channels and the r.m.s. noise (assuming typical ``shallow''
survey observations), and these values can be improved by one or more
orders of magnitude by employing full correlator capabilities and
dedicated targeted observations.

\bibliography{ms}

\begin{thebibliography}{75}%
\makeatletter
\providecommand \@ifxundefined [1]{%
 \@ifx{#1\undefined}
}%
\providecommand \@ifnum [1]{%
 \ifnum #1\expandafter \@firstoftwo
 \else \expandafter \@secondoftwo
 \fi
}%
\providecommand \@ifx [1]{%
 \ifx #1\expandafter \@firstoftwo
 \else \expandafter \@secondoftwo
 \fi
}%
\providecommand \natexlab [1]{#1}%
\providecommand \enquote  [1]{``#1''}%
\providecommand \bibnamefont  [1]{#1}%
\providecommand \bibfnamefont [1]{#1}%
\providecommand \citenamefont [1]{#1}%
\providecommand \href@noop [0]{\@secondoftwo}%
\providecommand \href [0]{\begingroup \@sanitize@url \@href}%
\providecommand \@href[1]{\@@startlink{#1}\@@href}%
\providecommand \@@href[1]{\endgroup#1\@@endlink}%
\providecommand \@sanitize@url [0]{\catcode `\\12\catcode `\$12\catcode
  `\&12\catcode `\#12\catcode `\^12\catcode `\_12\catcode `\%12\relax}%
\providecommand \@@startlink[1]{}%
\providecommand \@@endlink[0]{}%
\providecommand \url  [0]{\begingroup\@sanitize@url \@url }%
\providecommand \@url [1]{\endgroup\@href {#1}{\urlprefix }}%
\providecommand \urlprefix  [0]{URL }%
\providecommand \Eprint [0]{\href }%
\providecommand \doibase [0]{http://dx.doi.org/}%
\providecommand \selectlanguage [0]{\@gobble}%
\providecommand \bibinfo  [0]{\@secondoftwo}%
\providecommand \bibfield  [0]{\@secondoftwo}%
\providecommand \translation [1]{[#1]}%
\providecommand \BibitemOpen [0]{}%
\providecommand \bibitemStop [0]{}%
\providecommand \bibitemNoStop [0]{.\EOS\space}%
\providecommand \EOS [0]{\spacefactor3000\relax}%
\providecommand \BibitemShut  [1]{\csname bibitem#1\endcsname}%
\let\auto@bib@innerbib\@empty
\bibitem [{\citenamefont {{Jungman}}\ \emph {et~al.}(1996)\citenamefont
  {{Jungman}}, \citenamefont {{Kamionkowski}},\ and\ \citenamefont
  {{Griest}}}]{jungman1996}%
  \BibitemOpen
  \bibfield  {author} {\bibinfo {author} {\bibfnamefont {G.}~\bibnamefont
  {{Jungman}}}, \bibinfo {author} {\bibfnamefont {M.}~\bibnamefont
  {{Kamionkowski}}}, \ and\ \bibinfo {author} {\bibfnamefont {K.}~\bibnamefont
  {{Griest}}},\ }\href {\doibase 10.1016/0370-1573(95)00058-5} {\bibfield
  {journal} {\bibinfo  {journal} {\physrep}\ }\textbf {\bibinfo {volume}
  {267}},\ \bibinfo {pages} {195} (\bibinfo {year} {1996})},\ \Eprint
  {http://arxiv.org/abs/hep-ph/9506380} {hep-ph/9506380} \BibitemShut {NoStop}%
\bibitem [{\citenamefont {{Bertone}}\ \emph {et~al.}(2005)\citenamefont
  {{Bertone}}, \citenamefont {{Hooper}},\ and\ \citenamefont
  {{Silk}}}]{bertone2005}%
  \BibitemOpen
  \bibfield  {author} {\bibinfo {author} {\bibfnamefont {G.}~\bibnamefont
  {{Bertone}}}, \bibinfo {author} {\bibfnamefont {D.}~\bibnamefont {{Hooper}}},
  \ and\ \bibinfo {author} {\bibfnamefont {J.}~\bibnamefont {{Silk}}},\ }\href
  {\doibase 10.1016/j.physrep.2004.08.031} {\bibfield  {journal} {\bibinfo
  {journal} {\physrep}\ }\textbf {\bibinfo {volume} {405}},\ \bibinfo {pages}
  {279} (\bibinfo {year} {2005})},\ \Eprint
  {http://arxiv.org/abs/hep-ph/0404175} {hep-ph/0404175} \BibitemShut {NoStop}%
\bibitem [{\citenamefont {{Bertone}}(2010)}]{bertone2010}%
  \BibitemOpen
  \bibfield  {author} {\bibinfo {author} {\bibfnamefont {G.}~\bibnamefont
  {{Bertone}}},\ }\href {\doibase 10.1038/nature09509} {\bibfield  {journal}
  {\bibinfo  {journal} {\nat}\ }\textbf {\bibinfo {volume} {468}},\ \bibinfo
  {pages} {389} (\bibinfo {year} {2010})},\ \Eprint
  {http://arxiv.org/abs/1011.3532} {1011.3532} \BibitemShut {NoStop}%
\bibitem [{\citenamefont {{Okun}}(1982)}]{okun1982}%
  \BibitemOpen
  \bibfield  {author} {\bibinfo {author} {\bibfnamefont {L.~B.}\ \bibnamefont
  {{Okun}}},\ }\href@noop {} {\bibfield  {journal} {\bibinfo  {journal}
  {ZhETF}\ }\textbf {\bibinfo {volume} {83}},\ \bibinfo {pages} {892} (\bibinfo
  {year} {1982})}\BibitemShut {NoStop}%
\bibitem [{\citenamefont {{Georgi}}\ \emph {et~al.}(1984)\citenamefont
  {{Georgi}}, \citenamefont {{Ginsparg}},\ and\ \citenamefont
  {{Glashow}}}]{georgi1984}%
  \BibitemOpen
  \bibfield  {author} {\bibinfo {author} {\bibfnamefont {H.}~\bibnamefont
  {{Georgi}}}, \bibinfo {author} {\bibfnamefont {P.}~\bibnamefont
  {{Ginsparg}}}, \ and\ \bibinfo {author} {\bibfnamefont {S.~L.}\ \bibnamefont
  {{Glashow}}},\ }\href@noop {} {\bibfield  {journal} {\bibinfo  {journal}
  {\nat}\ }\textbf {\bibinfo {volume} {306}},\ \bibinfo {pages} {765} (\bibinfo
  {year} {1984})}\BibitemShut {NoStop}%
\bibitem [{\citenamefont {{Bergstr{\"o}m}}(2000)}]{bergstrom2000}%
  \BibitemOpen
  \bibfield  {author} {\bibinfo {author} {\bibfnamefont {L.}~\bibnamefont
  {{Bergstr{\"o}m}}},\ }\href {\doibase 10.1088/0034-4885/63/5/2r3} {\bibfield
  {journal} {\bibinfo  {journal} {Reports on Progress in Physics}\ }\textbf
  {\bibinfo {volume} {63}},\ \bibinfo {pages} {793} (\bibinfo {year} {2000})},\
  \Eprint {http://arxiv.org/abs/arXiv:hep-ph/0002126} {arXiv:hep-ph/0002126}
  \BibitemShut {NoStop}%
\bibitem [{\citenamefont {{Ahlers}}\ \emph {et~al.}(2008)\citenamefont
  {{Ahlers}}, \citenamefont {{Jaeckel}}, \citenamefont {{Redondo}},\ and\
  \citenamefont {{Ringwald}}}]{ahlers2008}%
  \BibitemOpen
  \bibfield  {author} {\bibinfo {author} {\bibfnamefont {M.}~\bibnamefont
  {{Ahlers}}}, \bibinfo {author} {\bibfnamefont {J.}~\bibnamefont {{Jaeckel}}},
  \bibinfo {author} {\bibfnamefont {J.}~\bibnamefont {{Redondo}}}, \ and\
  \bibinfo {author} {\bibfnamefont {A.}~\bibnamefont {{Ringwald}}},\ }\href
  {\doibase 10.1103/PhysRevD.78.075005} {\bibfield  {journal} {\bibinfo
  {journal} {Phys. Rev. D}\ }\textbf {\bibinfo {volume} {78}},\ \bibinfo {eid}
  {075005} (\bibinfo {year} {2008})},\ \Eprint {http://arxiv.org/abs/0807.4143}
  {arXiv:0807.4143 [hep-ph]} \BibitemShut {NoStop}%
\bibitem [{\citenamefont {{Jaeckel}}\ and\ \citenamefont
  {{Ringwald}}(2010)}]{jaeckel2010}%
  \BibitemOpen
  \bibfield  {author} {\bibinfo {author} {\bibfnamefont {J.}~\bibnamefont
  {{Jaeckel}}}\ and\ \bibinfo {author} {\bibfnamefont {A.}~\bibnamefont
  {{Ringwald}}},\ }\href {\doibase 10.1146/annurev.nucl.012809.104433}
  {\bibfield  {journal} {\bibinfo  {journal} {ARNPS}\ }\textbf {\bibinfo
  {volume} {60}},\ \bibinfo {pages} {405} (\bibinfo {year} {2010})},\ \Eprint
  {http://arxiv.org/abs/1002.0329} {arXiv:1002.0329 [hep-ph]} \BibitemShut
  {NoStop}%
\bibitem [{\citenamefont {{Ringwald}}(2012)}]{ringwald2012}%
  \BibitemOpen
  \bibfield  {author} {\bibinfo {author} {\bibfnamefont {A.}~\bibnamefont
  {{Ringwald}}},\ }\href@noop {} {\bibfield  {journal} {\bibinfo  {journal}
  {ArXiv e-prints}\ } (\bibinfo {year} {2012})},\ \Eprint
  {http://arxiv.org/abs/1210.5081} {arXiv:1210.5081 [hep-ph]} \BibitemShut
  {NoStop}%
\bibitem [{\citenamefont {{Jaeckel}}\ \emph {et~al.}(2007)\citenamefont
  {{Jaeckel}}, \citenamefont {{Mass{\'o}}}, \citenamefont {{Redondo}},
  \citenamefont {{Ringwald}},\ and\ \citenamefont {{Takahashi}}}]{jaeckel2007}%
  \BibitemOpen
  \bibfield  {author} {\bibinfo {author} {\bibfnamefont {J.}~\bibnamefont
  {{Jaeckel}}}, \bibinfo {author} {\bibfnamefont {E.}~\bibnamefont
  {{Mass{\'o}}}}, \bibinfo {author} {\bibfnamefont {J.}~\bibnamefont
  {{Redondo}}}, \bibinfo {author} {\bibfnamefont {A.}~\bibnamefont
  {{Ringwald}}}, \ and\ \bibinfo {author} {\bibfnamefont {F.}~\bibnamefont
  {{Takahashi}}},\ }\href {\doibase 10.1103/PhysRevD.75.013004} {\bibfield
  {journal} {\bibinfo  {journal} {Phys. Rev. D}\ }\textbf {\bibinfo {volume}
  {75}},\ \bibinfo {eid} {013004} (\bibinfo {year} {2007})},\ \Eprint
  {http://arxiv.org/abs/arXiv:hep-ph/0610203} {arXiv:hep-ph/0610203}
  \BibitemShut {NoStop}%
\bibitem [{\citenamefont {{Raffelt}}\ and\ \citenamefont
  {{Stodolsky}}(1988)}]{raffelt1988}%
  \BibitemOpen
  \bibfield  {author} {\bibinfo {author} {\bibfnamefont {G.}~\bibnamefont
  {{Raffelt}}}\ and\ \bibinfo {author} {\bibfnamefont {L.}~\bibnamefont
  {{Stodolsky}}},\ }\href {\doibase 10.1103/PhysRevD.37.1237} {\bibfield
  {journal} {\bibinfo  {journal} {Phys. Rev. D}\ }\textbf {\bibinfo {volume}
  {37}},\ \bibinfo {pages} {1237} (\bibinfo {year} {1988})}\BibitemShut
  {NoStop}%
\bibitem [{\citenamefont {{Zavattini}}\ \emph {et~al.}(2006)\citenamefont
  {{Zavattini}}, \citenamefont {{Zavattini}}, \citenamefont {{Ruoso}},
  \citenamefont {{Polacco}}, \citenamefont {{Milotti}}, \citenamefont
  {{Karuza}}, \citenamefont {{Gastaldi}}, \citenamefont {{di Domenico}},
  \citenamefont {{Della Valle}}, \citenamefont {{Cimino}}, \citenamefont
  {{Carusotto}}, \citenamefont {{Cantatore}},\ and\ \citenamefont
  {{Bregant}}}]{zavattini2006}%
  \BibitemOpen
  \bibfield  {author} {\bibinfo {author} {\bibfnamefont {E.}~\bibnamefont
  {{Zavattini}}}, \bibinfo {author} {\bibfnamefont {G.}~\bibnamefont
  {{Zavattini}}}, \bibinfo {author} {\bibfnamefont {G.}~\bibnamefont
  {{Ruoso}}}, \bibinfo {author} {\bibfnamefont {E.}~\bibnamefont {{Polacco}}},
  \bibinfo {author} {\bibfnamefont {E.}~\bibnamefont {{Milotti}}}, \bibinfo
  {author} {\bibfnamefont {M.}~\bibnamefont {{Karuza}}}, \bibinfo {author}
  {\bibfnamefont {U.}~\bibnamefont {{Gastaldi}}}, \bibinfo {author}
  {\bibfnamefont {G.}~\bibnamefont {{di Domenico}}}, \bibinfo {author}
  {\bibfnamefont {F.}~\bibnamefont {{Della Valle}}}, \bibinfo {author}
  {\bibfnamefont {R.}~\bibnamefont {{Cimino}}}, \bibinfo {author}
  {\bibfnamefont {S.}~\bibnamefont {{Carusotto}}}, \bibinfo {author}
  {\bibfnamefont {G.}~\bibnamefont {{Cantatore}}}, \ and\ \bibinfo {author}
  {\bibfnamefont {M.}~\bibnamefont {{Bregant}}},\ }\href {\doibase
  10.1103/PhysRevLett.96.110406} {\bibfield  {journal} {\bibinfo  {journal}
  {Phys. Rev. Lett.}\ }\textbf {\bibinfo {volume} {96}},\ \bibinfo {eid}
  {110406} (\bibinfo {year} {2006})},\ \Eprint
  {http://arxiv.org/abs/arXiv:hep-ex/0507107} {arXiv:hep-ex/0507107}
  \BibitemShut {NoStop}%
\bibitem [{\citenamefont {{Ahlers}}\ \emph {et~al.}(2007)\citenamefont
  {{Ahlers}}, \citenamefont {{Gies}}, \citenamefont {{Jaeckel}}, \citenamefont
  {{Redondo}},\ and\ \citenamefont {{Ringwald}}}]{ahlers2007}%
  \BibitemOpen
  \bibfield  {author} {\bibinfo {author} {\bibfnamefont {M.}~\bibnamefont
  {{Ahlers}}}, \bibinfo {author} {\bibfnamefont {H.}~\bibnamefont {{Gies}}},
  \bibinfo {author} {\bibfnamefont {J.}~\bibnamefont {{Jaeckel}}}, \bibinfo
  {author} {\bibfnamefont {J.}~\bibnamefont {{Redondo}}}, \ and\ \bibinfo
  {author} {\bibfnamefont {A.}~\bibnamefont {{Ringwald}}},\ }\href {\doibase
  10.1103/PhysRevD.76.115005} {\bibfield  {journal} {\bibinfo  {journal}
  {\prd}\ }\textbf {\bibinfo {volume} {76}},\ \bibinfo {eid} {115005} (\bibinfo
  {year} {2007})},\ \Eprint {http://arxiv.org/abs/0706.2836} {arXiv:0706.2836
  [hep-ph]} \BibitemShut {NoStop}%
\bibitem [{\citenamefont {{Redondo}}\ and\ \citenamefont
  {{Postma}}(2009)}]{redondo2009}%
  \BibitemOpen
  \bibfield  {author} {\bibinfo {author} {\bibfnamefont {J.}~\bibnamefont
  {{Redondo}}}\ and\ \bibinfo {author} {\bibfnamefont {M.}~\bibnamefont
  {{Postma}}},\ }\href {\doibase 10.1088/1475-7516/2009/02/005} {\bibfield
  {journal} {\bibinfo  {journal} {\jcap}\ }\textbf {\bibinfo {volume} {2}},\
  \bibinfo {eid} {005} (\bibinfo {year} {2009})},\ \Eprint
  {http://arxiv.org/abs/0811.0326} {arXiv:0811.0326 [hep-ph]} \BibitemShut
  {NoStop}%
\bibitem [{\citenamefont {{Arias}}\ \emph {et~al.}(2012)\citenamefont
  {{Arias}}, \citenamefont {{Cadamuro}}, \citenamefont {{Goodsell}},
  \citenamefont {{Jaeckel}}, \citenamefont {{Redondo}},\ and\ \citenamefont
  {{Ringwald}}}]{arias2012}%
  \BibitemOpen
  \bibfield  {author} {\bibinfo {author} {\bibfnamefont {P.}~\bibnamefont
  {{Arias}}}, \bibinfo {author} {\bibfnamefont {D.}~\bibnamefont {{Cadamuro}}},
  \bibinfo {author} {\bibfnamefont {M.}~\bibnamefont {{Goodsell}}}, \bibinfo
  {author} {\bibfnamefont {J.}~\bibnamefont {{Jaeckel}}}, \bibinfo {author}
  {\bibfnamefont {J.}~\bibnamefont {{Redondo}}}, \ and\ \bibinfo {author}
  {\bibfnamefont {A.}~\bibnamefont {{Ringwald}}},\ }\href {\doibase
  10.1088/1475-7516/2012/06/013} {\bibfield  {journal} {\bibinfo  {journal}
  {\jcap}\ }\textbf {\bibinfo {volume} {6}},\ \bibinfo {eid} {013} (\bibinfo
  {year} {2012})},\ \Eprint {http://arxiv.org/abs/1201.5902} {arXiv:1201.5902
  [hep-ph]} \BibitemShut {NoStop}%
\bibitem [{\citenamefont {{Holdom}}(1986)}]{holdom1986}%
  \BibitemOpen
  \bibfield  {author} {\bibinfo {author} {\bibfnamefont {B.}~\bibnamefont
  {{Holdom}}},\ }\href {\doibase 10.1016/0370-2693(86)91377-8} {\bibfield
  {journal} {\bibinfo  {journal} {Physics Letters B}\ }\textbf {\bibinfo
  {volume} {166}},\ \bibinfo {pages} {196} (\bibinfo {year}
  {1986})}\BibitemShut {NoStop}%
\bibitem [{\citenamefont {{Jaeckel}}\ \emph {et~al.}(2008)\citenamefont
  {{Jaeckel}}, \citenamefont {{Redondo}},\ and\ \citenamefont
  {{Ringwald}}}]{jaeckel2008}%
  \BibitemOpen
  \bibfield  {author} {\bibinfo {author} {\bibfnamefont {J.}~\bibnamefont
  {{Jaeckel}}}, \bibinfo {author} {\bibfnamefont {J.}~\bibnamefont
  {{Redondo}}}, \ and\ \bibinfo {author} {\bibfnamefont {A.}~\bibnamefont
  {{Ringwald}}},\ }\href {\doibase 10.1103/PhysRevLett.101.131801} {\bibfield
  {journal} {\bibinfo  {journal} {Phys. Rev. Lett.}\ }\textbf {\bibinfo
  {volume} {101}},\ \bibinfo {eid} {131801} (\bibinfo {year} {2008})},\ \Eprint
  {http://arxiv.org/abs/0804.4157} {arXiv:0804.4157} \BibitemShut {NoStop}%
\bibitem [{\citenamefont {{Redondo}}(2008)}]{redondo2008a}%
  \BibitemOpen
  \bibfield  {author} {\bibinfo {author} {\bibfnamefont {J.}~\bibnamefont
  {{Redondo}}},\ }\href {\doibase 10.1088/1475-7516/2008/07/008} {\bibfield
  {journal} {\bibinfo  {journal} {\jcap}\ }\textbf {\bibinfo {volume} {7}},\
  \bibinfo {eid} {008} (\bibinfo {year} {2008})},\ \Eprint
  {http://arxiv.org/abs/0801.1527} {arXiv:0801.1527 [hep-ph]} \BibitemShut
  {NoStop}%
\bibitem [{\citenamefont {{Dienes}}\ \emph {et~al.}(1997)\citenamefont
  {{Dienes}}, \citenamefont {{Kolda}},\ and\ \citenamefont
  {{March-Russell}}}]{dienes1997}%
  \BibitemOpen
  \bibfield  {author} {\bibinfo {author} {\bibfnamefont {K.~R.}\ \bibnamefont
  {{Dienes}}}, \bibinfo {author} {\bibfnamefont {C.}~\bibnamefont {{Kolda}}}, \
  and\ \bibinfo {author} {\bibfnamefont {J.}~\bibnamefont {{March-Russell}}},\
  }\href {\doibase 10.1016/S0550-3213(97)80028-4} {\bibfield  {journal}
  {\bibinfo  {journal} {Nuclear Physics B}\ }\textbf {\bibinfo {volume}
  {492}},\ \bibinfo {pages} {104} (\bibinfo {year} {1997})},\ \Eprint
  {http://arxiv.org/abs/arXiv:hep-ph/9610479} {arXiv:hep-ph/9610479}
  \BibitemShut {NoStop}%
\bibitem [{\citenamefont {{Abel}}\ \emph
  {et~al.}(2008{\natexlab{a}})\citenamefont {{Abel}}, \citenamefont
  {{Goodsell}}, \citenamefont {{Jaeckel}}, \citenamefont {{Khoze}},\ and\
  \citenamefont {{Ringwald}}}]{abel2008a}%
  \BibitemOpen
  \bibfield  {author} {\bibinfo {author} {\bibfnamefont {S.~A.}\ \bibnamefont
  {{Abel}}}, \bibinfo {author} {\bibfnamefont {M.~D.}\ \bibnamefont
  {{Goodsell}}}, \bibinfo {author} {\bibfnamefont {J.}~\bibnamefont
  {{Jaeckel}}}, \bibinfo {author} {\bibfnamefont {V.~V.}\ \bibnamefont
  {{Khoze}}}, \ and\ \bibinfo {author} {\bibfnamefont {A.}~\bibnamefont
  {{Ringwald}}},\ }\href {\doibase 10.1088/1126-6708/2008/07/124} {\bibfield
  {journal} {\bibinfo  {journal} {Journal of High Energy Physics}\ }\textbf
  {\bibinfo {volume} {7}},\ \bibinfo {pages} {124} (\bibinfo {year}
  {2008}{\natexlab{a}})},\ \Eprint {http://arxiv.org/abs/0803.1449}
  {arXiv:0803.1449 [hep-ph]} \BibitemShut {NoStop}%
\bibitem [{\citenamefont {{Abel}}\ \emph
  {et~al.}(2008{\natexlab{b}})\citenamefont {{Abel}}, \citenamefont
  {{Jaeckel}}, \citenamefont {{Khoze}},\ and\ \citenamefont
  {{Ringwald}}}]{abel2008b}%
  \BibitemOpen
  \bibfield  {author} {\bibinfo {author} {\bibfnamefont {S.~A.}\ \bibnamefont
  {{Abel}}}, \bibinfo {author} {\bibfnamefont {J.}~\bibnamefont {{Jaeckel}}},
  \bibinfo {author} {\bibfnamefont {V.~V.}\ \bibnamefont {{Khoze}}}, \ and\
  \bibinfo {author} {\bibfnamefont {A.}~\bibnamefont {{Ringwald}}},\ }\href
  {\doibase 10.1016/j.physletb.2008.03.076} {\bibfield  {journal} {\bibinfo
  {journal} {Physics Letters B}\ }\textbf {\bibinfo {volume} {666}},\ \bibinfo
  {pages} {66} (\bibinfo {year} {2008}{\natexlab{b}})},\ \Eprint
  {http://arxiv.org/abs/arXiv:hep-ph/0608248} {arXiv:hep-ph/0608248}
  \BibitemShut {NoStop}%
\bibitem [{\citenamefont {{Goodsell}}\ \emph {et~al.}(2009)\citenamefont
  {{Goodsell}}, \citenamefont {{Jaeckel}}, \citenamefont {{Redondo}},\ and\
  \citenamefont {{Ringwald}}}]{goodsell2009}%
  \BibitemOpen
  \bibfield  {author} {\bibinfo {author} {\bibfnamefont {M.}~\bibnamefont
  {{Goodsell}}}, \bibinfo {author} {\bibfnamefont {J.}~\bibnamefont
  {{Jaeckel}}}, \bibinfo {author} {\bibfnamefont {J.}~\bibnamefont
  {{Redondo}}}, \ and\ \bibinfo {author} {\bibfnamefont {A.}~\bibnamefont
  {{Ringwald}}},\ }\href {\doibase 10.1088/1126-6708/2009/11/027} {\bibfield
  {journal} {\bibinfo  {journal} {Journal of High Energy Physics}\ }\textbf
  {\bibinfo {volume} {11}},\ \bibinfo {eid} {027} (\bibinfo {year} {2009})},\
  \Eprint {http://arxiv.org/abs/0909.0515} {arXiv:0909.0515 [hep-ph]}
  \BibitemShut {NoStop}%
\bibitem [{\citenamefont {{Cicoli}}\ \emph {et~al.}(2011)\citenamefont
  {{Cicoli}}, \citenamefont {{Goodsell}}, \citenamefont {{Jaeckel}},\ and\
  \citenamefont {{Ringwald}}}]{cicoli2011}%
  \BibitemOpen
  \bibfield  {author} {\bibinfo {author} {\bibfnamefont {M.}~\bibnamefont
  {{Cicoli}}}, \bibinfo {author} {\bibfnamefont {M.}~\bibnamefont
  {{Goodsell}}}, \bibinfo {author} {\bibfnamefont {J.}~\bibnamefont
  {{Jaeckel}}}, \ and\ \bibinfo {author} {\bibfnamefont {A.}~\bibnamefont
  {{Ringwald}}},\ }\href {\doibase 10.1007/JHEP07(2011)114} {\bibfield
  {journal} {\bibinfo  {journal} {Journal of High Energy Physics}\ }\textbf
  {\bibinfo {volume} {7}},\ \bibinfo {pages} {114} (\bibinfo {year} {2011})},\
  \Eprint {http://arxiv.org/abs/1103.3705} {arXiv:1103.3705 [hep-th]}
  \BibitemShut {NoStop}%
\bibitem [{\citenamefont {{Asztalos}}\ \emph {et~al.}(2010)\citenamefont
  {{Asztalos}}, \citenamefont {{Carosi}}, \citenamefont {{Hagmann}},
  \citenamefont {{Kinion}}, \citenamefont {{van Bibber}}, \citenamefont
  {{Hotz}}, \citenamefont {{Rosenberg}}, \citenamefont {{Rybka}}, \citenamefont
  {{Hoskins}}, \citenamefont {{Hwang}}, \citenamefont {{Sikivie}},
  \citenamefont {{Tanner}}, \citenamefont {{Bradley}}, \citenamefont
  {{Clarke}},\ and\ \citenamefont {{ADMX Collaboration}}}]{asztalos2010}%
  \BibitemOpen
  \bibfield  {author} {\bibinfo {author} {\bibfnamefont {S.~J.}\ \bibnamefont
  {{Asztalos}}}, \bibinfo {author} {\bibfnamefont {G.}~\bibnamefont
  {{Carosi}}}, \bibinfo {author} {\bibfnamefont {C.}~\bibnamefont {{Hagmann}}},
  \bibinfo {author} {\bibfnamefont {D.}~\bibnamefont {{Kinion}}}, \bibinfo
  {author} {\bibfnamefont {K.}~\bibnamefont {{van Bibber}}}, \bibinfo {author}
  {\bibfnamefont {M.}~\bibnamefont {{Hotz}}}, \bibinfo {author} {\bibfnamefont
  {L.~J.}\ \bibnamefont {{Rosenberg}}}, \bibinfo {author} {\bibfnamefont
  {G.}~\bibnamefont {{Rybka}}}, \bibinfo {author} {\bibfnamefont
  {J.}~\bibnamefont {{Hoskins}}}, \bibinfo {author} {\bibfnamefont
  {J.}~\bibnamefont {{Hwang}}}, \bibinfo {author} {\bibfnamefont
  {P.}~\bibnamefont {{Sikivie}}}, \bibinfo {author} {\bibfnamefont {D.~B.}\
  \bibnamefont {{Tanner}}}, \bibinfo {author} {\bibfnamefont {R.}~\bibnamefont
  {{Bradley}}}, \bibinfo {author} {\bibfnamefont {J.}~\bibnamefont {{Clarke}}},
  \ and\ \bibinfo {author} {\bibnamefont {{ADMX Collaboration}}},\ }\href
  {\doibase 10.1103/PhysRevLett.104.041301} {\bibfield  {journal} {\bibinfo
  {journal} {Physical Review Letters}\ }\textbf {\bibinfo {volume} {104}},\
  \bibinfo {eid} {041301} (\bibinfo {year} {2010})},\ \Eprint
  {http://arxiv.org/abs/0910.5914} {arXiv:0910.5914 [astro-ph.CO]} \BibitemShut
  {NoStop}%
\bibitem [{\citenamefont {{Sikivie}}(2010)}]{sikivie2010}%
  \BibitemOpen
  \bibfield  {author} {\bibinfo {author} {\bibfnamefont {P.}~\bibnamefont
  {{Sikivie}}},\ }\href@noop {} {\bibfield  {journal} {\bibinfo  {journal}
  {ArXiv e-prints}\ } (\bibinfo {year} {2010})},\ \Eprint
  {http://arxiv.org/abs/1009.0762} {arXiv:1009.0762 [hep-ph]} \BibitemShut
  {NoStop}%
\bibitem [{\citenamefont {{Baker}}\ \emph {et~al.}(2012)\citenamefont
  {{Baker}}, \citenamefont {{Betz}}, \citenamefont {{Caspers}}, \citenamefont
  {{Jaeckel}}, \citenamefont {{Lindner}}, \citenamefont {{Ringwald}},
  \citenamefont {{Semertzidis}}, \citenamefont {{Sikivie}},\ and\ \citenamefont
  {{Zioutas}}}]{baker2012}%
  \BibitemOpen
  \bibfield  {author} {\bibinfo {author} {\bibfnamefont {O.~K.}\ \bibnamefont
  {{Baker}}}, \bibinfo {author} {\bibfnamefont {M.}~\bibnamefont {{Betz}}},
  \bibinfo {author} {\bibfnamefont {F.}~\bibnamefont {{Caspers}}}, \bibinfo
  {author} {\bibfnamefont {J.}~\bibnamefont {{Jaeckel}}}, \bibinfo {author}
  {\bibfnamefont {A.}~\bibnamefont {{Lindner}}}, \bibinfo {author}
  {\bibfnamefont {A.}~\bibnamefont {{Ringwald}}}, \bibinfo {author}
  {\bibfnamefont {Y.}~\bibnamefont {{Semertzidis}}}, \bibinfo {author}
  {\bibfnamefont {P.}~\bibnamefont {{Sikivie}}}, \ and\ \bibinfo {author}
  {\bibfnamefont {K.}~\bibnamefont {{Zioutas}}},\ }\href {\doibase
  10.1103/PhysRevD.85.035018} {\bibfield  {journal} {\bibinfo  {journal}
  {\prd}\ }\textbf {\bibinfo {volume} {85}},\ \bibinfo {eid} {035018} (\bibinfo
  {year} {2012})},\ \Eprint {http://arxiv.org/abs/1110.2180} {arXiv:1110.2180
  [physics.ins-det]} \BibitemShut {NoStop}%
\bibitem [{\citenamefont {{Afanasev}}\ \emph {et~al.}(2010)\citenamefont
  {{Afanasev}}, \citenamefont {{Baker}}, \citenamefont {{Beard}}, \citenamefont
  {{Biallas}}, \citenamefont {{Boyce}}, \citenamefont {{Minarni}},
  \citenamefont {{Ramdon}}, \citenamefont {{Shinn}},\ and\ \citenamefont
  {{Slocum}}}]{afanasiev2010}%
  \BibitemOpen
  \bibfield  {author} {\bibinfo {author} {\bibfnamefont {A.}~\bibnamefont
  {{Afanasev}}}, \bibinfo {author} {\bibfnamefont {O.~K.}\ \bibnamefont
  {{Baker}}}, \bibinfo {author} {\bibfnamefont {K.~B.}\ \bibnamefont
  {{Beard}}}, \bibinfo {author} {\bibfnamefont {G.}~\bibnamefont {{Biallas}}},
  \bibinfo {author} {\bibfnamefont {J.}~\bibnamefont {{Boyce}}}, \bibinfo
  {author} {\bibfnamefont {M.}~\bibnamefont {{Minarni}}}, \bibinfo {author}
  {\bibfnamefont {R.}~\bibnamefont {{Ramdon}}}, \bibinfo {author}
  {\bibfnamefont {M.}~\bibnamefont {{Shinn}}}, \ and\ \bibinfo {author}
  {\bibfnamefont {P.}~\bibnamefont {{Slocum}}},\ }in\ \href {\doibase
  10.1063/1.3327543} {\emph {\bibinfo {booktitle} {{American Institute of
  Physics Conference Series}}}},\ Vol.\ \bibinfo {volume} {1200},\ \bibinfo
  {editor} {edited by\ \bibinfo {editor} {\bibfnamefont {G.}~\bibnamefont
  {{Alverson}}}, \bibinfo {editor} {\bibfnamefont {P.}~\bibnamefont {{Nath}}},
  \ and\ \bibinfo {editor} {\bibfnamefont {B.}~\bibnamefont {{Nelson}}}}\
  (\bibinfo {year} {2010})\ pp.\ \bibinfo {pages} {1081--1084}\BibitemShut
  {NoStop}%
\bibitem [{\citenamefont {{Ehret}}\ \emph {et~al.}(2010)\citenamefont
  {{Ehret}}, \citenamefont {{Frede}}, \citenamefont {{Ghazaryan}},
  \citenamefont {{Hildebrandt}}, \citenamefont {{Knabbe}}, \citenamefont
  {{Kracht}}, \citenamefont {{Lindner}}, \citenamefont {{List}}, \citenamefont
  {{Meier}}, \citenamefont {{Meyer}}, \citenamefont {{Notz}}, \citenamefont
  {{Redondo}}, \citenamefont {{Ringwald}}, \citenamefont {{Wiedemann}},\ and\
  \citenamefont {{Willke}}}]{ehret2010}%
  \BibitemOpen
  \bibfield  {author} {\bibinfo {author} {\bibfnamefont {K.}~\bibnamefont
  {{Ehret}}}, \bibinfo {author} {\bibfnamefont {M.}~\bibnamefont {{Frede}}},
  \bibinfo {author} {\bibfnamefont {S.}~\bibnamefont {{Ghazaryan}}}, \bibinfo
  {author} {\bibfnamefont {M.}~\bibnamefont {{Hildebrandt}}}, \bibinfo {author}
  {\bibfnamefont {E.-A.}\ \bibnamefont {{Knabbe}}}, \bibinfo {author}
  {\bibfnamefont {D.}~\bibnamefont {{Kracht}}}, \bibinfo {author}
  {\bibfnamefont {A.}~\bibnamefont {{Lindner}}}, \bibinfo {author}
  {\bibfnamefont {J.}~\bibnamefont {{List}}}, \bibinfo {author} {\bibfnamefont
  {T.}~\bibnamefont {{Meier}}}, \bibinfo {author} {\bibfnamefont
  {N.}~\bibnamefont {{Meyer}}}, \bibinfo {author} {\bibfnamefont
  {D.}~\bibnamefont {{Notz}}}, \bibinfo {author} {\bibfnamefont
  {J.}~\bibnamefont {{Redondo}}}, \bibinfo {author} {\bibfnamefont
  {A.}~\bibnamefont {{Ringwald}}}, \bibinfo {author} {\bibfnamefont
  {G.}~\bibnamefont {{Wiedemann}}}, \ and\ \bibinfo {author} {\bibfnamefont
  {B.}~\bibnamefont {{Willke}}},\ }\href {\doibase
  10.1016/j.physletb.2010.04.066} {\bibfield  {journal} {\bibinfo  {journal}
  {Phys. Lett. B}\ }\textbf {\bibinfo {volume} {689}},\ \bibinfo {pages} {149}
  (\bibinfo {year} {2010})},\ \Eprint {http://arxiv.org/abs/1004.1313}
  {arXiv:1004.1313 [hep-ex]} \BibitemShut {NoStop}%
\bibitem [{\citenamefont {{Cadamuro}}\ and\ \citenamefont
  {{Redondo}}(2010)}]{cadamuro2010}%
  \BibitemOpen
  \bibfield  {author} {\bibinfo {author} {\bibfnamefont {D.}~\bibnamefont
  {{Cadamuro}}}\ and\ \bibinfo {author} {\bibfnamefont {J.}~\bibnamefont
  {{Redondo}}},\ }\href@noop {} {\bibfield  {journal} {\bibinfo  {journal}
  {ArXiv:hep-ph/1010.4689}\ } (\bibinfo {year} {2010})},\ \Eprint
  {http://arxiv.org/abs/1010.4689} {arXiv:1010.4689 [hep-ph]} \BibitemShut
  {NoStop}%
\bibitem [{\citenamefont {{Wagner}}\ \emph {et~al.}(2010)\citenamefont
  {{Wagner}}, \citenamefont {{Rybka}}, \citenamefont {{Hotz}}, \citenamefont
  {{Rosenberg}}, \citenamefont {{Asztalos}}, \citenamefont {{Carosi}},
  \citenamefont {{Hagmann}}, \citenamefont {{Kinion}}, \citenamefont {{van
  Bibber}}, \citenamefont {{Hoskins}}, \citenamefont {{Martin}}, \citenamefont
  {{Sikivie}}, \citenamefont {{Tanner}}, \citenamefont {{Bradley}},\ and\
  \citenamefont {{Clarke}}}]{wagner2010}%
  \BibitemOpen
  \bibfield  {author} {\bibinfo {author} {\bibfnamefont {A.}~\bibnamefont
  {{Wagner}}}, \bibinfo {author} {\bibfnamefont {G.}~\bibnamefont {{Rybka}}},
  \bibinfo {author} {\bibfnamefont {M.}~\bibnamefont {{Hotz}}}, \bibinfo
  {author} {\bibfnamefont {L.~J.}\ \bibnamefont {{Rosenberg}}}, \bibinfo
  {author} {\bibfnamefont {S.~J.}\ \bibnamefont {{Asztalos}}}, \bibinfo
  {author} {\bibfnamefont {G.}~\bibnamefont {{Carosi}}}, \bibinfo {author}
  {\bibfnamefont {C.}~\bibnamefont {{Hagmann}}}, \bibinfo {author}
  {\bibfnamefont {D.}~\bibnamefont {{Kinion}}}, \bibinfo {author}
  {\bibfnamefont {K.}~\bibnamefont {{van Bibber}}}, \bibinfo {author}
  {\bibfnamefont {J.}~\bibnamefont {{Hoskins}}}, \bibinfo {author}
  {\bibfnamefont {C.}~\bibnamefont {{Martin}}}, \bibinfo {author}
  {\bibfnamefont {P.}~\bibnamefont {{Sikivie}}}, \bibinfo {author}
  {\bibfnamefont {D.~B.}\ \bibnamefont {{Tanner}}}, \bibinfo {author}
  {\bibfnamefont {R.}~\bibnamefont {{Bradley}}}, \ and\ \bibinfo {author}
  {\bibfnamefont {J.}~\bibnamefont {{Clarke}}},\ }\href {\doibase
  10.1103/PhysRevLett.105.171801} {\bibfield  {journal} {\bibinfo  {journal}
  {Physical Review Letters}\ }\textbf {\bibinfo {volume} {105}},\ \bibinfo
  {eid} {171801} (\bibinfo {year} {2010})},\ \Eprint
  {http://arxiv.org/abs/1007.3766} {arXiv:1007.3766 [hep-ex]} \BibitemShut
  {NoStop}%
\bibitem [{\citenamefont {{Andreas}}\ \emph {et~al.}(2012)\citenamefont
  {{Andreas}}, \citenamefont {{Niebuhr}},\ and\ \citenamefont
  {{Ringwald}}}]{andreas2012}%
  \BibitemOpen
  \bibfield  {author} {\bibinfo {author} {\bibfnamefont {S.}~\bibnamefont
  {{Andreas}}}, \bibinfo {author} {\bibfnamefont {C.}~\bibnamefont
  {{Niebuhr}}}, \ and\ \bibinfo {author} {\bibfnamefont {A.}~\bibnamefont
  {{Ringwald}}},\ }\href@noop {} {\bibfield  {journal} {\bibinfo  {journal}
  {ArXiv e-prints}\ } (\bibinfo {year} {2012})},\ \Eprint
  {http://arxiv.org/abs/1209.6083} {arXiv:1209.6083 [hep-ph]} \BibitemShut
  {NoStop}%
\bibitem [{\citenamefont {{Betz}}\ and\ \citenamefont
  {{Caspers}}(2012)}]{betz2012}%
  \BibitemOpen
  \bibfield  {author} {\bibinfo {author} {\bibfnamefont {M.}~\bibnamefont
  {{Betz}}}\ and\ \bibinfo {author} {\bibfnamefont {F.}~\bibnamefont
  {{Caspers}}},\ }\href@noop {} {\bibfield  {journal} {\bibinfo  {journal}
  {ArXiv e-prints}\ } (\bibinfo {year} {2012})},\ \Eprint
  {http://arxiv.org/abs/1207.3275} {arXiv:1207.3275 [physics.ins-det]}
  \BibitemShut {NoStop}%
\bibitem [{\citenamefont {{Zechlin}}\ \emph {et~al.}(2008)\citenamefont
  {{Zechlin}}, \citenamefont {{Horns}},\ and\ \citenamefont
  {{Redondo}}}]{zechlin2008}%
  \BibitemOpen
  \bibfield  {author} {\bibinfo {author} {\bibfnamefont {H.-S.}\ \bibnamefont
  {{Zechlin}}}, \bibinfo {author} {\bibfnamefont {D.}~\bibnamefont {{Horns}}},
  \ and\ \bibinfo {author} {\bibfnamefont {J.}~\bibnamefont {{Redondo}}},\ }in\
  \href {\doibase 10.1063/1.3076781} {\emph {\bibinfo {booktitle} {American
  Institute of Physics Conference Series}}},\ Vol.\ \bibinfo {volume} {1085},\
  \bibinfo {editor} {edited by\ \bibinfo {editor} {\bibnamefont
  {{F.~A.~Aharonian, W.~Hofmann, \& F.~Rieger}}}}\ (\bibinfo {year} {2008})\
  pp.\ \bibinfo {pages} {727--730},\ \Eprint {http://arxiv.org/abs/0810.5501}
  {arXiv:0810.5501} \BibitemShut {NoStop}%
\bibitem [{\citenamefont {{Mirizzi}}\ \emph {et~al.}(2009)\citenamefont
  {{Mirizzi}}, \citenamefont {{Redondo}},\ and\ \citenamefont
  {{Sigl}}}]{mirizzi2009a}%
  \BibitemOpen
  \bibfield  {author} {\bibinfo {author} {\bibfnamefont {A.}~\bibnamefont
  {{Mirizzi}}}, \bibinfo {author} {\bibfnamefont {J.}~\bibnamefont
  {{Redondo}}}, \ and\ \bibinfo {author} {\bibfnamefont {G.}~\bibnamefont
  {{Sigl}}},\ }\href {\doibase 10.1088/1475-7516/2009/03/026} {\bibfield
  {journal} {\bibinfo  {journal} {JCAP}\ }\textbf {\bibinfo {volume} {3}},\
  \bibinfo {pages} {26} (\bibinfo {year} {2009})},\ \Eprint
  {http://arxiv.org/abs/0901.0014} {arXiv:0901.0014 [hep-ph]} \BibitemShut
  {NoStop}%
\bibitem [{\citenamefont {{Arias}}\ \emph {et~al.}(2010)\citenamefont
  {{Arias}}, \citenamefont {{Jaeckel}}, \citenamefont {{Redondo}},\ and\
  \citenamefont {{Ringwald}}}]{arias2010}%
  \BibitemOpen
  \bibfield  {author} {\bibinfo {author} {\bibfnamefont {P.}~\bibnamefont
  {{Arias}}}, \bibinfo {author} {\bibfnamefont {J.}~\bibnamefont {{Jaeckel}}},
  \bibinfo {author} {\bibfnamefont {J.}~\bibnamefont {{Redondo}}}, \ and\
  \bibinfo {author} {\bibfnamefont {A.}~\bibnamefont {{Ringwald}}},\ }\href
  {\doibase 10.1103/PhysRevD.82.115018} {\bibfield  {journal} {\bibinfo
  {journal} {Phys. Rev. D}\ }\textbf {\bibinfo {volume} {82}},\ \bibinfo {eid}
  {115018} (\bibinfo {year} {2010})},\ \Eprint {http://arxiv.org/abs/1009.4875}
  {arXiv:1009.4875 [hep-ph]} \BibitemShut {NoStop}%
\bibitem [{\citenamefont {{Redondo}}(2010)}]{redondo2010a}%
  \BibitemOpen
  \bibfield  {author} {\bibinfo {author} {\bibfnamefont {J.}~\bibnamefont
  {{Redondo}}},\ }\href@noop {} {\bibfield  {journal} {\bibinfo  {journal}
  {ArXiv:hep-ph/1002.0447}\ } (\bibinfo {year} {2010})},\ \Eprint
  {http://arxiv.org/abs/1002.0447} {arXiv:1002.0447 [hep-ph]} \BibitemShut
  {NoStop}%
\bibitem [{\citenamefont {{Redondo}}\ and\ \citenamefont
  {{Ringwald}}(2011)}]{redondo2011}%
  \BibitemOpen
  \bibfield  {author} {\bibinfo {author} {\bibfnamefont {J.}~\bibnamefont
  {{Redondo}}}\ and\ \bibinfo {author} {\bibfnamefont {A.}~\bibnamefont
  {{Ringwald}}},\ }\href {\doibase 10.1080/00107514.2011.563516} {\bibfield
  {journal} {\bibinfo  {journal} {Contemporary Physics}\ }\textbf {\bibinfo
  {volume} {52}},\ \bibinfo {pages} {211} (\bibinfo {year} {2011})},\ \Eprint
  {http://arxiv.org/abs/1011.3741} {arXiv:1011.3741 [hep-ph]} \BibitemShut
  {NoStop}%
\bibitem [{\citenamefont {{Schwarz}}\ \emph {et~al.}(2012)\citenamefont
  {{Schwarz}}, \citenamefont {{Lindner}}, \citenamefont {{Redondo}},
  \citenamefont {{Ringwald}},\ and\ \citenamefont {{Wiedemann}}}]{schwarz2011}%
  \BibitemOpen
  \bibfield  {author} {\bibinfo {author} {\bibfnamefont {M.}~\bibnamefont
  {{Schwarz}}}, \bibinfo {author} {\bibfnamefont {A.}~\bibnamefont
  {{Lindner}}}, \bibinfo {author} {\bibfnamefont {J.}~\bibnamefont
  {{Redondo}}}, \bibinfo {author} {\bibfnamefont {A.}~\bibnamefont
  {{Ringwald}}}, \ and\ \bibinfo {author} {\bibfnamefont {G.}~\bibnamefont
  {{Wiedemann}}},\ }\href@noop {} {\bibfield  {journal} {\bibinfo  {journal}
  {ArXiv e-prints}\ } (\bibinfo {year} {2012})},\ \Eprint
  {http://arxiv.org/abs/1111.5797} {arXiv:1111.5797 [astro-ph.IM]} \BibitemShut
  {NoStop}%
\bibitem [{\citenamefont {{Goldhaber}}\ and\ \citenamefont
  {{Nieto}}(1971)}]{goldhaber1971}%
  \BibitemOpen
  \bibfield  {author} {\bibinfo {author} {\bibfnamefont {A.~S.}\ \bibnamefont
  {{Goldhaber}}}\ and\ \bibinfo {author} {\bibfnamefont {M.~M.}\ \bibnamefont
  {{Nieto}}},\ }\href {\doibase 10.1103/RevModPhys.43.277} {\bibfield
  {journal} {\bibinfo  {journal} {Reviews of Modern Physics}\ }\textbf
  {\bibinfo {volume} {43}},\ \bibinfo {pages} {277} (\bibinfo {year}
  {1971})}\BibitemShut {NoStop}%
\bibitem [{Note1()}]{Note1}%
  \BibitemOpen
  \bibinfo {note} {1\protect \,Jy = $10^{-26}$\protect \,J\protect
  \,m$^{-2}$\protect \,s$^{-1}$\protect \,Hz$^{-1}$}\BibitemShut {NoStop}%
\bibitem [{Note2()}]{Note2}%
  \BibitemOpen
  \bibinfo {note} {Square Kilometer Array, a next generation radio telescope
  that will provide about a two orders of magnitude improvement in imaging
  sensitivity and surveying speed for radio observations in the 0.3--20\protect
  \,GHz frequency range; {\protect \tt
  http://www.skatelescope.org}.}\BibitemShut {Stop}%
\bibitem [{Note3()}]{Note3}%
  \BibitemOpen
  \bibinfo {note} {{\protect \tt http://www.ska.ac.za/meerkat}}\BibitemShut
  {NoStop}%
\bibitem [{Note4()}]{Note4}%
  \BibitemOpen
  \bibinfo {note} {{\protect \tt http://www.atnf.csiro.au/SKA}}\BibitemShut
  {NoStop}%
\bibitem [{\citenamefont {{Baars}}\ \emph {et~al.}(1977)\citenamefont
  {{Baars}}, \citenamefont {{Genzel}}, \citenamefont {{Pauliny-Toth}},\ and\
  \citenamefont {{Witzel}}}]{baars1977}%
  \BibitemOpen
  \bibfield  {author} {\bibinfo {author} {\bibfnamefont {J.~W.~M.}\
  \bibnamefont {{Baars}}}, \bibinfo {author} {\bibfnamefont {R.}~\bibnamefont
  {{Genzel}}}, \bibinfo {author} {\bibfnamefont {I.~I.~K.}\ \bibnamefont
  {{Pauliny-Toth}}}, \ and\ \bibinfo {author} {\bibfnamefont {A.}~\bibnamefont
  {{Witzel}}},\ }\href@noop {} {\bibfield  {journal} {\bibinfo  {journal}
  {A\&A}\ }\textbf {\bibinfo {volume} {61}},\ \bibinfo {pages} {99} (\bibinfo
  {year} {1977})}\BibitemShut {NoStop}%
\bibitem [{\citenamefont {{Kuo}}\ and\ \citenamefont
  {{Pantaleone}}(1989)}]{kuo1989}%
  \BibitemOpen
  \bibfield  {author} {\bibinfo {author} {\bibfnamefont {T.~K.}\ \bibnamefont
  {{Kuo}}}\ and\ \bibinfo {author} {\bibfnamefont {J.}~\bibnamefont
  {{Pantaleone}}},\ }\href {\doibase 10.1103/PhysRevD.39.1930} {\bibfield
  {journal} {\bibinfo  {journal} {\prd}\ }\textbf {\bibinfo {volume} {39}},\
  \bibinfo {pages} {1930} (\bibinfo {year} {1989})}\BibitemShut {NoStop}%
\bibitem [{\citenamefont {{Nussinov}}(1976)}]{nussinov1976}%
  \BibitemOpen
  \bibfield  {author} {\bibinfo {author} {\bibfnamefont {S.}~\bibnamefont
  {{Nussinov}}},\ }\href {\doibase 10.1016/0370-2693(76)90648-1} {\bibfield
  {journal} {\bibinfo  {journal} {Physics Letters B}\ }\textbf {\bibinfo
  {volume} {63}},\ \bibinfo {pages} {201} (\bibinfo {year} {1976})}\BibitemShut
  {NoStop}%
\bibitem [{\citenamefont {{Giunti}}\ and\ \citenamefont
  {{Kim}}(1998)}]{giunti1998}%
  \BibitemOpen
  \bibfield  {author} {\bibinfo {author} {\bibfnamefont {C.}~\bibnamefont
  {{Giunti}}}\ and\ \bibinfo {author} {\bibfnamefont {C.~W.}\ \bibnamefont
  {{Kim}}},\ }\href {\doibase 10.1103/PhysRevD.58.017301} {\bibfield  {journal}
  {\bibinfo  {journal} {\prd}\ }\textbf {\bibinfo {volume} {58}},\ \bibinfo
  {eid} {017301} (\bibinfo {year} {1998})},\ \Eprint
  {http://arxiv.org/abs/arXiv:hep-ph/9711363} {arXiv:hep-ph/9711363}
  \BibitemShut {NoStop}%
\bibitem [{\citenamefont {{Meyer}}(2010)}]{meyer2010}%
  \BibitemOpen
  \bibfield  {author} {\bibinfo {author} {\bibfnamefont {M.}~\bibnamefont
  {{Meyer}}},\ }\href@noop {} {\bibinfo {type} {Diploma thesis}},\ \bibinfo
  {school} {University of Hamburg} (\bibinfo {year} {2010})\BibitemShut
  {NoStop}%
\bibitem [{\citenamefont {{Hinshaw}}\ \emph {et~al.}(2009)\citenamefont
  {{Hinshaw}}, \citenamefont {{Weiland}}, \citenamefont {{Hill}}, \citenamefont
  {{Odegard}}, \citenamefont {{Larson}}, \citenamefont {{Bennett}},
  \citenamefont {{Dunkley}}, \citenamefont {{Gold}}, \citenamefont {{Greason}},
  \citenamefont {{Jarosik}}, \citenamefont {{Komatsu}}, \citenamefont
  {{Nolta}}, \citenamefont {{Page}}, \citenamefont {{Spergel}}, \citenamefont
  {{Wollack}}, \citenamefont {{Halpern}}, \citenamefont {{Kogut}},
  \citenamefont {{Limon}}, \citenamefont {{Meyer}}, \citenamefont {{Tucker}},\
  and\ \citenamefont {{Wright}}}]{hinshaw2009}%
  \BibitemOpen
  \bibfield  {author} {\bibinfo {author} {\bibfnamefont {G.}~\bibnamefont
  {{Hinshaw}}}, \bibinfo {author} {\bibfnamefont {J.~L.}\ \bibnamefont
  {{Weiland}}}, \bibinfo {author} {\bibfnamefont {R.~S.}\ \bibnamefont
  {{Hill}}}, \bibinfo {author} {\bibfnamefont {N.}~\bibnamefont {{Odegard}}},
  \bibinfo {author} {\bibfnamefont {D.}~\bibnamefont {{Larson}}}, \bibinfo
  {author} {\bibfnamefont {C.~L.}\ \bibnamefont {{Bennett}}}, \bibinfo {author}
  {\bibfnamefont {J.}~\bibnamefont {{Dunkley}}}, \bibinfo {author}
  {\bibfnamefont {B.}~\bibnamefont {{Gold}}}, \bibinfo {author} {\bibfnamefont
  {M.~R.}\ \bibnamefont {{Greason}}}, \bibinfo {author} {\bibfnamefont
  {N.}~\bibnamefont {{Jarosik}}}, \bibinfo {author} {\bibfnamefont
  {E.}~\bibnamefont {{Komatsu}}}, \bibinfo {author} {\bibfnamefont {M.~R.}\
  \bibnamefont {{Nolta}}}, \bibinfo {author} {\bibfnamefont {L.}~\bibnamefont
  {{Page}}}, \bibinfo {author} {\bibfnamefont {D.~N.}\ \bibnamefont
  {{Spergel}}}, \bibinfo {author} {\bibfnamefont {E.}~\bibnamefont
  {{Wollack}}}, \bibinfo {author} {\bibfnamefont {M.}~\bibnamefont
  {{Halpern}}}, \bibinfo {author} {\bibfnamefont {A.}~\bibnamefont {{Kogut}}},
  \bibinfo {author} {\bibfnamefont {M.}~\bibnamefont {{Limon}}}, \bibinfo
  {author} {\bibfnamefont {S.~S.}\ \bibnamefont {{Meyer}}}, \bibinfo {author}
  {\bibfnamefont {G.~S.}\ \bibnamefont {{Tucker}}}, \ and\ \bibinfo {author}
  {\bibfnamefont {E.~L.}\ \bibnamefont {{Wright}}},\ }\href {\doibase
  10.1088/0067-0049/180/2/225} {\bibfield  {journal} {\bibinfo  {journal}
  {\apjs}\ }\textbf {\bibinfo {volume} {180}},\ \bibinfo {pages} {225}
  (\bibinfo {year} {2009})},\ \Eprint {http://arxiv.org/abs/0803.0732}
  {arXiv:0803.0732} \BibitemShut {NoStop}%
\bibitem [{\citenamefont {{Gottl{\"o}ber}}\ \emph {et~al.}(2003)\citenamefont
  {{Gottl{\"o}ber}}, \citenamefont {{{\L}okas}}, \citenamefont {{Klypin}},\
  and\ \citenamefont {{Hoffman}}}]{gottloeber2003}%
  \BibitemOpen
  \bibfield  {author} {\bibinfo {author} {\bibfnamefont {S.}~\bibnamefont
  {{Gottl{\"o}ber}}}, \bibinfo {author} {\bibfnamefont {E.~L.}\ \bibnamefont
  {{{\L}okas}}}, \bibinfo {author} {\bibfnamefont {A.}~\bibnamefont
  {{Klypin}}}, \ and\ \bibinfo {author} {\bibfnamefont {Y.}~\bibnamefont
  {{Hoffman}}},\ }\href {\doibase 10.1046/j.1365-8711.2003.06850.x} {\bibfield
  {journal} {\bibinfo  {journal} {\mnras}\ }\textbf {\bibinfo {volume} {344}},\
  \bibinfo {pages} {715} (\bibinfo {year} {2003})},\ \Eprint
  {http://arxiv.org/abs/arXiv:astro-ph/0305393} {arXiv:astro-ph/0305393}
  \BibitemShut {NoStop}%
\bibitem [{\citenamefont {{Pynzar'}}\ and\ \citenamefont
  {{Shishov}}(2008)}]{pynzar2008}%
  \BibitemOpen
  \bibfield  {author} {\bibinfo {author} {\bibfnamefont {A.~V.}\ \bibnamefont
  {{Pynzar'}}}\ and\ \bibinfo {author} {\bibfnamefont {V.~I.}\ \bibnamefont
  {{Shishov}}},\ }\href {\doibase 10.1134/S1063772908080039} {\bibfield
  {journal} {\bibinfo  {journal} {Astronomy Reports}\ }\textbf {\bibinfo
  {volume} {52}},\ \bibinfo {pages} {623} (\bibinfo {year} {2008})}\BibitemShut
  {NoStop}%
\bibitem [{\citenamefont {{Taylor}}\ and\ \citenamefont
  {{Cordes}}(1993)}]{taylor1993}%
  \BibitemOpen
  \bibfield  {author} {\bibinfo {author} {\bibfnamefont {J.~H.}\ \bibnamefont
  {{Taylor}}}\ and\ \bibinfo {author} {\bibfnamefont {J.~M.}\ \bibnamefont
  {{Cordes}}},\ }\href {\doibase 10.1086/172870} {\bibfield  {journal}
  {\bibinfo  {journal} {\apj}\ }\textbf {\bibinfo {volume} {411}},\ \bibinfo
  {pages} {674} (\bibinfo {year} {1993})}\BibitemShut {NoStop}%
\bibitem [{\citenamefont {{Cordes}}\ and\ \citenamefont
  {{Lazio}}(2002)}]{cordes2002}%
  \BibitemOpen
  \bibfield  {author} {\bibinfo {author} {\bibfnamefont {J.~M.}\ \bibnamefont
  {{Cordes}}}\ and\ \bibinfo {author} {\bibfnamefont {T.~J.~W.}\ \bibnamefont
  {{Lazio}}},\ }\href@noop {} {\bibfield  {journal} {\bibinfo  {journal}
  {ArXiv:astro-ph/0207156}\ } (\bibinfo {year} {2002})},\ \Eprint
  {http://arxiv.org/abs/arXiv:astro-ph/0207156} {arXiv:astro-ph/0207156}
  \BibitemShut {NoStop}%
\bibitem [{\citenamefont {{Plionis}}\ and\ \citenamefont
  {{Basilakos}}(2002)}]{plionis2002}%
  \BibitemOpen
  \bibfield  {author} {\bibinfo {author} {\bibfnamefont {M.}~\bibnamefont
  {{Plionis}}}\ and\ \bibinfo {author} {\bibfnamefont {S.}~\bibnamefont
  {{Basilakos}}},\ }\href {\doibase 10.1046/j.1365-8711.2002.05069.x}
  {\bibfield  {journal} {\bibinfo  {journal} {\mnras}\ }\textbf {\bibinfo
  {volume} {330}},\ \bibinfo {pages} {399} (\bibinfo {year} {2002})},\ \Eprint
  {http://arxiv.org/abs/arXiv:astro-ph/0106491} {arXiv:astro-ph/0106491}
  \BibitemShut {NoStop}%
\bibitem [{\citenamefont {{Pan}}\ \emph {et~al.}(2012)\citenamefont {{Pan}},
  \citenamefont {{Vogeley}}, \citenamefont {{Hoyle}}, \citenamefont {{Choi}},\
  and\ \citenamefont {{Park}}}]{pan2012}%
  \BibitemOpen
  \bibfield  {author} {\bibinfo {author} {\bibfnamefont {D.~C.}\ \bibnamefont
  {{Pan}}}, \bibinfo {author} {\bibfnamefont {M.~S.}\ \bibnamefont
  {{Vogeley}}}, \bibinfo {author} {\bibfnamefont {F.}~\bibnamefont {{Hoyle}}},
  \bibinfo {author} {\bibfnamefont {Y.-Y.}\ \bibnamefont {{Choi}}}, \ and\
  \bibinfo {author} {\bibfnamefont {C.}~\bibnamefont {{Park}}},\ }\href
  {\doibase 10.1111/j.1365-2966.2011.20197.x} {\bibfield  {journal} {\bibinfo
  {journal} {\mnras}\ }\textbf {\bibinfo {volume} {421}},\ \bibinfo {pages}
  {926} (\bibinfo {year} {2012})},\ \Eprint {http://arxiv.org/abs/1103.4156}
  {arXiv:1103.4156 [astro-ph.CO]} \BibitemShut {NoStop}%
\bibitem [{\citenamefont {{Kravtsov}}\ \emph {et~al.}(2002)\citenamefont
  {{Kravtsov}}, \citenamefont {{Klypin}},\ and\ \citenamefont
  {{Hoffman}}}]{kravtsov2002}%
  \BibitemOpen
  \bibfield  {author} {\bibinfo {author} {\bibfnamefont {A.~V.}\ \bibnamefont
  {{Kravtsov}}}, \bibinfo {author} {\bibfnamefont {A.}~\bibnamefont
  {{Klypin}}}, \ and\ \bibinfo {author} {\bibfnamefont {Y.}~\bibnamefont
  {{Hoffman}}},\ }\href {\doibase 10.1086/340046} {\bibfield  {journal}
  {\bibinfo  {journal} {\apj}\ }\textbf {\bibinfo {volume} {571}},\ \bibinfo
  {pages} {563} (\bibinfo {year} {2002})},\ \Eprint
  {http://arxiv.org/abs/arXiv:astro-ph/0109077} {arXiv:astro-ph/0109077}
  \BibitemShut {NoStop}%
\bibitem [{\citenamefont {{Cen}}\ \emph {et~al.}(2005)\citenamefont {{Cen}},
  \citenamefont {{Nagamine}},\ and\ \citenamefont {{Ostriker}}}]{cen2005}%
  \BibitemOpen
  \bibfield  {author} {\bibinfo {author} {\bibfnamefont {R.}~\bibnamefont
  {{Cen}}}, \bibinfo {author} {\bibfnamefont {K.}~\bibnamefont {{Nagamine}}}, \
  and\ \bibinfo {author} {\bibfnamefont {J.~P.}\ \bibnamefont {{Ostriker}}},\
  }\href {\doibase 10.1086/497353} {\bibfield  {journal} {\bibinfo  {journal}
  {\apj}\ }\textbf {\bibinfo {volume} {635}},\ \bibinfo {pages} {86} (\bibinfo
  {year} {2005})},\ \Eprint {http://arxiv.org/abs/arXiv:astro-ph/0407143}
  {arXiv:astro-ph/0407143} \BibitemShut {NoStop}%
\bibitem [{\citenamefont {{Komatsu}}\ \emph {et~al.}(2011)\citenamefont
  {{Komatsu}}, \citenamefont {{Smith}}, \citenamefont {{Dunkley}},
  \citenamefont {{Bennett}}, \citenamefont {{Gold}}, \citenamefont {{Hinshaw}},
  \citenamefont {{Jarosik}}, \citenamefont {{Larson}}, \citenamefont {{Nolta}},
  \citenamefont {{Page}}, \citenamefont {{Spergel}}, \citenamefont {{Halpern}},
  \citenamefont {{Hill}}, \citenamefont {{Kogut}}, \citenamefont {{Limon}},
  \citenamefont {{Meyer}}, \citenamefont {{Odegard}}, \citenamefont {{Tucker}},
  \citenamefont {{Weiland}}, \citenamefont {{Wollack}},\ and\ \citenamefont
  {{Wright}}}]{komatsu2011}%
  \BibitemOpen
  \bibfield  {author} {\bibinfo {author} {\bibfnamefont {E.}~\bibnamefont
  {{Komatsu}}}, \bibinfo {author} {\bibfnamefont {K.~M.}\ \bibnamefont
  {{Smith}}}, \bibinfo {author} {\bibfnamefont {J.}~\bibnamefont {{Dunkley}}},
  \bibinfo {author} {\bibfnamefont {C.~L.}\ \bibnamefont {{Bennett}}}, \bibinfo
  {author} {\bibfnamefont {B.}~\bibnamefont {{Gold}}}, \bibinfo {author}
  {\bibfnamefont {G.}~\bibnamefont {{Hinshaw}}}, \bibinfo {author}
  {\bibfnamefont {N.}~\bibnamefont {{Jarosik}}}, \bibinfo {author}
  {\bibfnamefont {D.}~\bibnamefont {{Larson}}}, \bibinfo {author}
  {\bibfnamefont {M.~R.}\ \bibnamefont {{Nolta}}}, \bibinfo {author}
  {\bibfnamefont {L.}~\bibnamefont {{Page}}}, \bibinfo {author} {\bibfnamefont
  {D.~N.}\ \bibnamefont {{Spergel}}}, \bibinfo {author} {\bibfnamefont
  {M.}~\bibnamefont {{Halpern}}}, \bibinfo {author} {\bibfnamefont {R.~S.}\
  \bibnamefont {{Hill}}}, \bibinfo {author} {\bibfnamefont {A.}~\bibnamefont
  {{Kogut}}}, \bibinfo {author} {\bibfnamefont {M.}~\bibnamefont {{Limon}}},
  \bibinfo {author} {\bibfnamefont {S.~S.}\ \bibnamefont {{Meyer}}}, \bibinfo
  {author} {\bibfnamefont {N.}~\bibnamefont {{Odegard}}}, \bibinfo {author}
  {\bibfnamefont {G.~S.}\ \bibnamefont {{Tucker}}}, \bibinfo {author}
  {\bibfnamefont {J.~L.}\ \bibnamefont {{Weiland}}}, \bibinfo {author}
  {\bibfnamefont {E.}~\bibnamefont {{Wollack}}}, \ and\ \bibinfo {author}
  {\bibfnamefont {E.~L.}\ \bibnamefont {{Wright}}},\ }\href {\doibase
  10.1088/0067-0049/192/2/18} {\bibfield  {journal} {\bibinfo  {journal}
  {\apjs}\ }\textbf {\bibinfo {volume} {192}},\ \bibinfo {eid} {18} (\bibinfo
  {year} {2011})},\ \Eprint {http://arxiv.org/abs/1001.4538} {arXiv:1001.4538
  [astro-ph.CO]} \BibitemShut {NoStop}%
\bibitem [{\citenamefont {{Sheth}}\ and\ \citenamefont {{van de
  Weygaert}}(2004)}]{sheth2004}%
  \BibitemOpen
  \bibfield  {author} {\bibinfo {author} {\bibfnamefont {R.~K.}\ \bibnamefont
  {{Sheth}}}\ and\ \bibinfo {author} {\bibfnamefont {R.}~\bibnamefont {{van de
  Weygaert}}},\ }\href {\doibase 10.1111/j.1365-2966.2004.07661.x} {\bibfield
  {journal} {\bibinfo  {journal} {\mnras}\ }\textbf {\bibinfo {volume} {350}},\
  \bibinfo {pages} {517} (\bibinfo {year} {2004})},\ \Eprint
  {http://arxiv.org/abs/arXiv:astro-ph/0311260} {arXiv:astro-ph/0311260}
  \BibitemShut {NoStop}%
\bibitem [{\citenamefont {{Fesen}}\ \emph {et~al.}(2006)\citenamefont
  {{Fesen}}, \citenamefont {{Hammell}}, \citenamefont {{Morse}}, \citenamefont
  {{Chevalier}}, \citenamefont {{Borkowski}}, \citenamefont {{Dopita}},
  \citenamefont {{Gerardy}}, \citenamefont {{Lawrence}}, \citenamefont
  {{Raymond}},\ and\ \citenamefont {{van den Bergh}}}]{fesen2006}%
  \BibitemOpen
  \bibfield  {author} {\bibinfo {author} {\bibfnamefont {R.~A.}\ \bibnamefont
  {{Fesen}}}, \bibinfo {author} {\bibfnamefont {M.~C.}\ \bibnamefont
  {{Hammell}}}, \bibinfo {author} {\bibfnamefont {J.}~\bibnamefont {{Morse}}},
  \bibinfo {author} {\bibfnamefont {R.~A.}\ \bibnamefont {{Chevalier}}},
  \bibinfo {author} {\bibfnamefont {K.~J.}\ \bibnamefont {{Borkowski}}},
  \bibinfo {author} {\bibfnamefont {M.~A.}\ \bibnamefont {{Dopita}}}, \bibinfo
  {author} {\bibfnamefont {C.~L.}\ \bibnamefont {{Gerardy}}}, \bibinfo {author}
  {\bibfnamefont {S.~S.}\ \bibnamefont {{Lawrence}}}, \bibinfo {author}
  {\bibfnamefont {J.~C.}\ \bibnamefont {{Raymond}}}, \ and\ \bibinfo {author}
  {\bibfnamefont {S.}~\bibnamefont {{van den Bergh}}},\ }\href {\doibase
  10.1086/504254} {\bibfield  {journal} {\bibinfo  {journal} {\apj}\ }\textbf
  {\bibinfo {volume} {645}},\ \bibinfo {pages} {283} (\bibinfo {year}
  {2006})},\ \Eprint {http://arxiv.org/abs/arXiv:astro-ph/0603371}
  {arXiv:astro-ph/0603371} \BibitemShut {NoStop}%
\bibitem [{\citenamefont {{Weiland}}\ \emph {et~al.}(2011)\citenamefont
  {{Weiland}}, \citenamefont {{Odegard}}, \citenamefont {{Hill}}, \citenamefont
  {{Wollack}}, \citenamefont {{Hinshaw}}, \citenamefont {{Greason}},
  \citenamefont {{Jarosik}}, \citenamefont {{Page}}, \citenamefont {{Bennett}},
  \citenamefont {{Dunkley}}, \citenamefont {{Gold}}, \citenamefont {{Halpern}},
  \citenamefont {{Kogut}}, \citenamefont {{Komatsu}}, \citenamefont {{Larson}},
  \citenamefont {{Limon}}, \citenamefont {{Meyer}}, \citenamefont {{Nolta}},
  \citenamefont {{Smith}}, \citenamefont {{Spergel}}, \citenamefont
  {{Tucker}},\ and\ \citenamefont {{Wright}}}]{weiland2011}%
  \BibitemOpen
  \bibfield  {author} {\bibinfo {author} {\bibfnamefont {J.~L.}\ \bibnamefont
  {{Weiland}}}, \bibinfo {author} {\bibfnamefont {N.}~\bibnamefont
  {{Odegard}}}, \bibinfo {author} {\bibfnamefont {R.~S.}\ \bibnamefont
  {{Hill}}}, \bibinfo {author} {\bibfnamefont {E.}~\bibnamefont {{Wollack}}},
  \bibinfo {author} {\bibfnamefont {G.}~\bibnamefont {{Hinshaw}}}, \bibinfo
  {author} {\bibfnamefont {M.~R.}\ \bibnamefont {{Greason}}}, \bibinfo {author}
  {\bibfnamefont {N.}~\bibnamefont {{Jarosik}}}, \bibinfo {author}
  {\bibfnamefont {L.}~\bibnamefont {{Page}}}, \bibinfo {author} {\bibfnamefont
  {C.~L.}\ \bibnamefont {{Bennett}}}, \bibinfo {author} {\bibfnamefont
  {J.}~\bibnamefont {{Dunkley}}}, \bibinfo {author} {\bibfnamefont
  {B.}~\bibnamefont {{Gold}}}, \bibinfo {author} {\bibfnamefont
  {M.}~\bibnamefont {{Halpern}}}, \bibinfo {author} {\bibfnamefont
  {A.}~\bibnamefont {{Kogut}}}, \bibinfo {author} {\bibfnamefont
  {E.}~\bibnamefont {{Komatsu}}}, \bibinfo {author} {\bibfnamefont
  {D.}~\bibnamefont {{Larson}}}, \bibinfo {author} {\bibfnamefont
  {M.}~\bibnamefont {{Limon}}}, \bibinfo {author} {\bibfnamefont {S.~S.}\
  \bibnamefont {{Meyer}}}, \bibinfo {author} {\bibfnamefont {M.~R.}\
  \bibnamefont {{Nolta}}}, \bibinfo {author} {\bibfnamefont {K.~M.}\
  \bibnamefont {{Smith}}}, \bibinfo {author} {\bibfnamefont {D.~N.}\
  \bibnamefont {{Spergel}}}, \bibinfo {author} {\bibfnamefont {G.~S.}\
  \bibnamefont {{Tucker}}}, \ and\ \bibinfo {author} {\bibfnamefont {E.~L.}\
  \bibnamefont {{Wright}}},\ }\href {\doibase 10.1088/0067-0049/192/2/19}
  {\bibfield  {journal} {\bibinfo  {journal} {\apjs}\ }\textbf {\bibinfo
  {volume} {192}},\ \bibinfo {eid} {19} (\bibinfo {year} {2011})},\ \Eprint
  {http://arxiv.org/abs/1001.4731} {arXiv:1001.4731 [astro-ph.CO]} \BibitemShut
  {NoStop}%
\bibitem [{\citenamefont {{Baars}}(1972)}]{baars1972}%
  \BibitemOpen
  \bibfield  {author} {\bibinfo {author} {\bibfnamefont {J.~W.~M.}\
  \bibnamefont {{Baars}}},\ }\href@noop {} {\bibfield  {journal} {\bibinfo
  {journal} {\aap}\ }\textbf {\bibinfo {volume} {17}},\ \bibinfo {pages} {172}
  (\bibinfo {year} {1972})}\BibitemShut {NoStop}%
\bibitem [{\citenamefont {{Winkel}}\ \emph {et~al.}(2012)\citenamefont
  {{Winkel}}, \citenamefont {{Kraus}},\ and\ \citenamefont
  {{Bach}}}]{winkel2012}%
  \BibitemOpen
  \bibfield  {author} {\bibinfo {author} {\bibfnamefont {B.}~\bibnamefont
  {{Winkel}}}, \bibinfo {author} {\bibfnamefont {A.}~\bibnamefont {{Kraus}}}, \
  and\ \bibinfo {author} {\bibfnamefont {U.}~\bibnamefont {{Bach}}},\ }\href
  {\doibase 10.1051/0004-6361/201118092} {\bibfield  {journal} {\bibinfo
  {journal} {\aap}\ }\textbf {\bibinfo {volume} {540}},\ \bibinfo {eid} {A140}
  (\bibinfo {year} {2012})},\ \Eprint {http://arxiv.org/abs/1203.0741}
  {arXiv:1203.0741 [astro-ph.IM]} \BibitemShut {NoStop}%
\bibitem [{\citenamefont {{Tappin}}(1986)}]{tappin1986}%
  \BibitemOpen
  \bibfield  {author} {\bibinfo {author} {\bibfnamefont {S.~J.}\ \bibnamefont
  {{Tappin}}},\ }\href {\doibase 10.1016/0032-0633(86)90106-6} {\bibfield
  {journal} {\bibinfo  {journal} {\planss}\ }\textbf {\bibinfo {volume} {34}},\
  \bibinfo {pages} {93} (\bibinfo {year} {1986})}\BibitemShut {NoStop}%
\bibitem [{\citenamefont {{Leahy}}\ \emph
  {et~al.}(1983{\natexlab{a}})\citenamefont {{Leahy}}, \citenamefont
  {{Darbro}}, \citenamefont {{Elsner}}, \citenamefont {{Weisskopf}},
  \citenamefont {{Kahn}}, \citenamefont {{Sutherland}},\ and\ \citenamefont
  {{Grindlay}}}]{leahy1983a}%
  \BibitemOpen
  \bibfield  {author} {\bibinfo {author} {\bibfnamefont {D.~A.}\ \bibnamefont
  {{Leahy}}}, \bibinfo {author} {\bibfnamefont {W.}~\bibnamefont {{Darbro}}},
  \bibinfo {author} {\bibfnamefont {R.~F.}\ \bibnamefont {{Elsner}}}, \bibinfo
  {author} {\bibfnamefont {M.~C.}\ \bibnamefont {{Weisskopf}}}, \bibinfo
  {author} {\bibfnamefont {S.}~\bibnamefont {{Kahn}}}, \bibinfo {author}
  {\bibfnamefont {P.~G.}\ \bibnamefont {{Sutherland}}}, \ and\ \bibinfo
  {author} {\bibfnamefont {J.~E.}\ \bibnamefont {{Grindlay}}},\ }\href
  {\doibase 10.1086/160766} {\bibfield  {journal} {\bibinfo  {journal} {ApJ}\
  }\textbf {\bibinfo {volume} {266}},\ \bibinfo {pages} {160} (\bibinfo {year}
  {1983}{\natexlab{a}})}\BibitemShut {NoStop}%
\bibitem [{\citenamefont {{Brazier}}(1994)}]{brazier1994}%
  \BibitemOpen
  \bibfield  {author} {\bibinfo {author} {\bibfnamefont {K.~T.~S.}\
  \bibnamefont {{Brazier}}},\ }\href@noop {} {\bibfield  {journal} {\bibinfo
  {journal} {MNRAS}\ }\textbf {\bibinfo {volume} {268}},\ \bibinfo {pages}
  {709} (\bibinfo {year} {1994})}\BibitemShut {NoStop}%
\bibitem [{\citenamefont {{Buccheri}}\ \emph {et~al.}(1983)\citenamefont
  {{Buccheri}}, \citenamefont {{Bennett}}, \citenamefont {{Bignami}},
  \citenamefont {{Bloemen}}, \citenamefont {{Boriakoff}}, \citenamefont
  {{Caraveo}}, \citenamefont {{Hermsen}}, \citenamefont {{Kanbach}},
  \citenamefont {{Manchester}}, \citenamefont {{Masnou}}, \citenamefont
  {{Mayer-Hasselwander}}, \citenamefont {{Ozel}}, \citenamefont {{Paul}},
  \citenamefont {{Sacco}}, \citenamefont {{Scarsi}},\ and\ \citenamefont
  {{Strong}}}]{buccheri1983}%
  \BibitemOpen
  \bibfield  {author} {\bibinfo {author} {\bibfnamefont {R.}~\bibnamefont
  {{Buccheri}}}, \bibinfo {author} {\bibfnamefont {K.}~\bibnamefont
  {{Bennett}}}, \bibinfo {author} {\bibfnamefont {G.~F.}\ \bibnamefont
  {{Bignami}}}, \bibinfo {author} {\bibfnamefont {J.~B.~G.~M.}\ \bibnamefont
  {{Bloemen}}}, \bibinfo {author} {\bibfnamefont {V.}~\bibnamefont
  {{Boriakoff}}}, \bibinfo {author} {\bibfnamefont {P.~A.}\ \bibnamefont
  {{Caraveo}}}, \bibinfo {author} {\bibfnamefont {W.}~\bibnamefont
  {{Hermsen}}}, \bibinfo {author} {\bibfnamefont {G.}~\bibnamefont
  {{Kanbach}}}, \bibinfo {author} {\bibfnamefont {R.~N.}\ \bibnamefont
  {{Manchester}}}, \bibinfo {author} {\bibfnamefont {J.~L.}\ \bibnamefont
  {{Masnou}}}, \bibinfo {author} {\bibfnamefont {H.~A.}\ \bibnamefont
  {{Mayer-Hasselwander}}}, \bibinfo {author} {\bibfnamefont {M.~E.}\
  \bibnamefont {{Ozel}}}, \bibinfo {author} {\bibfnamefont {J.~A.}\
  \bibnamefont {{Paul}}}, \bibinfo {author} {\bibfnamefont {B.}~\bibnamefont
  {{Sacco}}}, \bibinfo {author} {\bibfnamefont {L.}~\bibnamefont {{Scarsi}}}, \
  and\ \bibinfo {author} {\bibfnamefont {A.~W.}\ \bibnamefont {{Strong}}},\
  }\href@noop {} {\bibfield  {journal} {\bibinfo  {journal} {A\&A}\ }\textbf
  {\bibinfo {volume} {128}},\ \bibinfo {pages} {245} (\bibinfo {year}
  {1983})}\BibitemShut {NoStop}%
\bibitem [{\citenamefont {{Caucci}}\ \emph {et~al.}(2008)\citenamefont
  {{Caucci}}, \citenamefont {{Colombi}}, \citenamefont {{Pichon}},
  \citenamefont {{Rollinde}}, \citenamefont {{Petitjean}},\ and\ \citenamefont
  {{Sousbie}}}]{caucci2008}%
  \BibitemOpen
  \bibfield  {author} {\bibinfo {author} {\bibfnamefont {S.}~\bibnamefont
  {{Caucci}}}, \bibinfo {author} {\bibfnamefont {S.}~\bibnamefont {{Colombi}}},
  \bibinfo {author} {\bibfnamefont {C.}~\bibnamefont {{Pichon}}}, \bibinfo
  {author} {\bibfnamefont {E.}~\bibnamefont {{Rollinde}}}, \bibinfo {author}
  {\bibfnamefont {P.}~\bibnamefont {{Petitjean}}}, \ and\ \bibinfo {author}
  {\bibfnamefont {T.}~\bibnamefont {{Sousbie}}},\ }\href {\doibase
  10.1111/j.1365-2966.2008.13016.x} {\bibfield  {journal} {\bibinfo  {journal}
  {\mnras}\ }\textbf {\bibinfo {volume} {386}},\ \bibinfo {pages} {211}
  (\bibinfo {year} {2008})}\BibitemShut {NoStop}%
\bibitem [{\citenamefont {{Abdalla}}\ and\ \citenamefont
  {{Rawlings}}(2005)}]{abdalla2005}%
  \BibitemOpen
  \bibfield  {author} {\bibinfo {author} {\bibfnamefont {F.~B.}\ \bibnamefont
  {{Abdalla}}}\ and\ \bibinfo {author} {\bibfnamefont {S.}~\bibnamefont
  {{Rawlings}}},\ }\href {\doibase 10.1111/j.1365-2966.2005.08650.x} {\bibfield
   {journal} {\bibinfo  {journal} {\mnras}\ }\textbf {\bibinfo {volume}
  {360}},\ \bibinfo {pages} {27} (\bibinfo {year} {2005})},\ \Eprint
  {http://arxiv.org/abs/arXiv:astro-ph/0411342} {arXiv:astro-ph/0411342}
  \BibitemShut {NoStop}%
\bibitem [{\citenamefont {{Duffy}}\ \emph {et~al.}(2008)\citenamefont
  {{Duffy}}, \citenamefont {{Battye}}, \citenamefont {{Davies}}, \citenamefont
  {{Moss}},\ and\ \citenamefont {{Wilkinson}}}]{duffy2008}%
  \BibitemOpen
  \bibfield  {author} {\bibinfo {author} {\bibfnamefont {A.~R.}\ \bibnamefont
  {{Duffy}}}, \bibinfo {author} {\bibfnamefont {R.~A.}\ \bibnamefont
  {{Battye}}}, \bibinfo {author} {\bibfnamefont {R.~D.}\ \bibnamefont
  {{Davies}}}, \bibinfo {author} {\bibfnamefont {A.}~\bibnamefont {{Moss}}}, \
  and\ \bibinfo {author} {\bibfnamefont {P.~N.}\ \bibnamefont {{Wilkinson}}},\
  }\href {\doibase 10.1111/j.1365-2966.2007.12537.x} {\bibfield  {journal}
  {\bibinfo  {journal} {\mnras}\ }\textbf {\bibinfo {volume} {383}},\ \bibinfo
  {pages} {150} (\bibinfo {year} {2008})},\ \Eprint
  {http://arxiv.org/abs/0707.2316} {arXiv:0707.2316} \BibitemShut {NoStop}%
\bibitem [{\citenamefont {{R{\"o}ttgering}}\ \emph {et~al.}(2011)\citenamefont
  {{R{\"o}ttgering}}, \citenamefont {{Afonso}}, \citenamefont {{Barthel}},
  \citenamefont {{Batejat}}, \citenamefont {{Best}}, \citenamefont
  {{Bonafede}}, \citenamefont {{Br{\"u}ggen}}, \citenamefont {{Brunetti}},
  \citenamefont {{Chy{\.z}y}}, \citenamefont {{Conway}}, \citenamefont
  {{Gasperin}}, \citenamefont {{Ferrari}}, \citenamefont {{Haverkorn}},
  \citenamefont {{Heald}}, \citenamefont {{Hoeft}}, \citenamefont {{Jackson}},
  \citenamefont {{Jarvis}}, \citenamefont {{Ker}}, \citenamefont {{Lehnert}},
  \citenamefont {{Macario}}, \citenamefont {{McKean}}, \citenamefont {{Miley}},
  \citenamefont {{Morganti}}, \citenamefont {{Oosterloo}}, \citenamefont
  {{Orr{\`u}}}, \citenamefont {{Pizzo}}, \citenamefont {{Rafferty}},
  \citenamefont {{Shulevski}}, \citenamefont {{Tasse}}, \citenamefont
  {{Bemmel}}, \citenamefont {{van der Tol}}, \citenamefont {{van Weeren}},
  \citenamefont {{Verheijen}}, \citenamefont {{White}},\ and\ \citenamefont
  {{Wise}}}]{rottgering2011}%
  \BibitemOpen
  \bibfield  {author} {\bibinfo {author} {\bibfnamefont {H.}~\bibnamefont
  {{R{\"o}ttgering}}}, \bibinfo {author} {\bibfnamefont {J.}~\bibnamefont
  {{Afonso}}}, \bibinfo {author} {\bibfnamefont {P.}~\bibnamefont {{Barthel}}},
  \bibinfo {author} {\bibfnamefont {F.}~\bibnamefont {{Batejat}}}, \bibinfo
  {author} {\bibfnamefont {P.}~\bibnamefont {{Best}}}, \bibinfo {author}
  {\bibfnamefont {A.}~\bibnamefont {{Bonafede}}}, \bibinfo {author}
  {\bibfnamefont {M.}~\bibnamefont {{Br{\"u}ggen}}}, \bibinfo {author}
  {\bibfnamefont {G.}~\bibnamefont {{Brunetti}}}, \bibinfo {author}
  {\bibfnamefont {K.}~\bibnamefont {{Chy{\.z}y}}}, \bibinfo {author}
  {\bibfnamefont {J.}~\bibnamefont {{Conway}}}, \bibinfo {author}
  {\bibfnamefont {F.~D.}\ \bibnamefont {{Gasperin}}}, \bibinfo {author}
  {\bibfnamefont {C.}~\bibnamefont {{Ferrari}}}, \bibinfo {author}
  {\bibfnamefont {M.}~\bibnamefont {{Haverkorn}}}, \bibinfo {author}
  {\bibfnamefont {G.}~\bibnamefont {{Heald}}}, \bibinfo {author} {\bibfnamefont
  {M.}~\bibnamefont {{Hoeft}}}, \bibinfo {author} {\bibfnamefont
  {N.}~\bibnamefont {{Jackson}}}, \bibinfo {author} {\bibfnamefont
  {M.}~\bibnamefont {{Jarvis}}}, \bibinfo {author} {\bibfnamefont
  {L.}~\bibnamefont {{Ker}}}, \bibinfo {author} {\bibfnamefont
  {M.}~\bibnamefont {{Lehnert}}}, \bibinfo {author} {\bibfnamefont
  {G.}~\bibnamefont {{Macario}}}, \bibinfo {author} {\bibfnamefont
  {J.}~\bibnamefont {{McKean}}}, \bibinfo {author} {\bibfnamefont
  {G.}~\bibnamefont {{Miley}}}, \bibinfo {author} {\bibfnamefont
  {R.}~\bibnamefont {{Morganti}}}, \bibinfo {author} {\bibfnamefont
  {T.}~\bibnamefont {{Oosterloo}}}, \bibinfo {author} {\bibfnamefont
  {E.}~\bibnamefont {{Orr{\`u}}}}, \bibinfo {author} {\bibfnamefont
  {R.}~\bibnamefont {{Pizzo}}}, \bibinfo {author} {\bibfnamefont
  {D.}~\bibnamefont {{Rafferty}}}, \bibinfo {author} {\bibfnamefont
  {A.}~\bibnamefont {{Shulevski}}}, \bibinfo {author} {\bibfnamefont
  {C.}~\bibnamefont {{Tasse}}}, \bibinfo {author} {\bibfnamefont {I.~V.}\
  \bibnamefont {{Bemmel}}}, \bibinfo {author} {\bibfnamefont {B.}~\bibnamefont
  {{van der Tol}}}, \bibinfo {author} {\bibfnamefont {R.}~\bibnamefont {{van
  Weeren}}}, \bibinfo {author} {\bibfnamefont {M.}~\bibnamefont {{Verheijen}}},
  \bibinfo {author} {\bibfnamefont {G.}~\bibnamefont {{White}}}, \ and\
  \bibinfo {author} {\bibfnamefont {M.}~\bibnamefont {{Wise}}},\ }\href
  {\doibase 10.1007/s12036-011-9129-x} {\bibfield  {journal} {\bibinfo
  {journal} {Journal of Astrophysics and Astronomy}\ }\textbf {\bibinfo
  {volume} {32}},\ \bibinfo {pages} {557} (\bibinfo {year} {2011})},\ \Eprint
  {http://arxiv.org/abs/1107.1606} {arXiv:1107.1606 [astro-ph.CO]} \BibitemShut
  {NoStop}%
\bibitem [{\citenamefont {{Duffy}}\ \emph {et~al.}(2012)\citenamefont
  {{Duffy}}, \citenamefont {{Meyer}}, \citenamefont {{Staveley-Smith}},
  \citenamefont {{Bernyk}}, \citenamefont {{Croton}}, \citenamefont
  {{Koribalski}}, \citenamefont {{Gerstmann}},\ and\ \citenamefont
  {{Westerlund}}}]{duffy2012}%
  \BibitemOpen
  \bibfield  {author} {\bibinfo {author} {\bibfnamefont {A.~R.}\ \bibnamefont
  {{Duffy}}}, \bibinfo {author} {\bibfnamefont {M.~J.}\ \bibnamefont
  {{Meyer}}}, \bibinfo {author} {\bibfnamefont {L.}~\bibnamefont
  {{Staveley-Smith}}}, \bibinfo {author} {\bibfnamefont {M.}~\bibnamefont
  {{Bernyk}}}, \bibinfo {author} {\bibfnamefont {D.~J.}\ \bibnamefont
  {{Croton}}}, \bibinfo {author} {\bibfnamefont {B.~S.}\ \bibnamefont
  {{Koribalski}}}, \bibinfo {author} {\bibfnamefont {D.}~\bibnamefont
  {{Gerstmann}}}, \ and\ \bibinfo {author} {\bibfnamefont {S.}~\bibnamefont
  {{Westerlund}}},\ }\href {\doibase 10.1111/j.1365-2966.2012.21987.x}
  {\bibfield  {journal} {\bibinfo  {journal} {\mnras}\ }\textbf {\bibinfo
  {volume} {426}},\ \bibinfo {pages} {3385} (\bibinfo {year} {2012})},\ \Eprint
  {http://arxiv.org/abs/1208.5592} {arXiv:1208.5592 [astro-ph.CO]} \BibitemShut
  {NoStop}%
\bibitem [{\citenamefont {{Leahy}}\ \emph
  {et~al.}(1983{\natexlab{b}})\citenamefont {{Leahy}}, \citenamefont
  {{Elsner}},\ and\ \citenamefont {{Weisskopf}}}]{leahy1983b}%
  \BibitemOpen
  \bibfield  {author} {\bibinfo {author} {\bibfnamefont {D.~A.}\ \bibnamefont
  {{Leahy}}}, \bibinfo {author} {\bibfnamefont {R.~F.}\ \bibnamefont
  {{Elsner}}}, \ and\ \bibinfo {author} {\bibfnamefont {M.~C.}\ \bibnamefont
  {{Weisskopf}}},\ }\href {\doibase 10.1086/161288} {\bibfield  {journal}
  {\bibinfo  {journal} {ApJ}\ }\textbf {\bibinfo {volume} {272}},\ \bibinfo
  {pages} {256} (\bibinfo {year} {1983}{\natexlab{b}})}\BibitemShut {NoStop}%
\bibitem [{\citenamefont {{Mardia}}\ and\ \citenamefont
  {{Jupp}}(2000)}]{mardia2000}%
  \BibitemOpen
  \bibfield  {author} {\bibinfo {author} {\bibfnamefont {K.~V.}\ \bibnamefont
  {{Mardia}}}\ and\ \bibinfo {author} {\bibfnamefont {P.~E.}\ \bibnamefont
  {{Jupp}}},\ }\href@noop {} {\emph {\bibinfo {title} {{Directional
  Statistics}}}}\ (\bibinfo  {publisher} {Wiley \& Sons: Chichester},\ \bibinfo
  {year} {2000})\BibitemShut {NoStop}%
\bibitem [{\citenamefont {{de Jager}}\ \emph {et~al.}(1989)\citenamefont {{de
  Jager}}, \citenamefont {{Raubenheimer}},\ and\ \citenamefont
  {{Swanepoel}}}]{dejager1989}%
  \BibitemOpen
  \bibfield  {author} {\bibinfo {author} {\bibfnamefont {O.~C.}\ \bibnamefont
  {{de Jager}}}, \bibinfo {author} {\bibfnamefont {B.~C.}\ \bibnamefont
  {{Raubenheimer}}}, \ and\ \bibinfo {author} {\bibfnamefont {J.~W.~H.}\
  \bibnamefont {{Swanepoel}}},\ }\href@noop {} {\bibfield  {journal} {\bibinfo
  {journal} {A\&A}\ }\textbf {\bibinfo {volume} {221}},\ \bibinfo {pages} {180}
  (\bibinfo {year} {1989})}\BibitemShut {NoStop}%
\end{thebibliography}%

\end{document}